\documentclass[aps,showpacs,showkeys]{revtex4} 
\usepackage{amsfonts}
\usepackage{amsmath}
\usepackage{amssymb}
\usepackage{color}
\usepackage{graphicx}

\begin{document}

\title{Exact results for nonequilibrium dynamics in Wigner phase space}
\author{K. Bencheikh}
\email{bencheikhkml@univ-setif.dz}
\affiliation{D\'{e}partement de Physique. Laboratoire de physique quantique et syst\`{e}%
mes dynamiques. Universit\'{e} Ferhat Abbas S\'{e}tif-1, Setif 19000, Algeria}
\author{L. M. Nieto}
\email{luismiguel.nieto.calzada@uva.es}
\affiliation{Departamento de F\'{\i}sica Te\'{o}rica, At\'{o}mica y \'{O}ptica and IMUVA,
Universidad de Valladolid, 47011 Valladolid, Spain}
\date{\today }

\begin{abstract}
We study time evolution of Wigner function of an initially interacting one-dimensional quantum gas following the switch-off of the interactions. For the scenario where at $t=0$ the interactions are suddenly suppressed, we derive a relationship between the dynamical Wigner function and its initial value. A two-particle system initially interacting through two different interactions of Dirac delta type is examined. For a system of particles that is suddenly let to move ballistically (without interactions) in a harmonic trap in d dimensions, and using time evolution of one-body density matrix, we derive a relationship between the time dependent Wigner function and its initial value. Using the inverse Wigner transform we obtain, for an initially harmonically trapped noninteracting particles in $d$ dimensions, the scaling law satisfied by the density matrix at time $t$ after a sudden change of the trapping frequency. Finally, the effects of interactions are analyzed in the dynamical Wigner function.
\end{abstract}

\maketitle

\section{Introduction}
\label{introduction}

The description of time evolution properties of quantum many body systems is
presently a fundamental topic of research. The amazing development of
trapping and cooling techniques have led to experimental realization of
quantum many-body systems consisting of ultracold atomic gases where atoms
are confined in traps \cite{Giorgini,Bloch}. These artificial many-body
systems can be produced and loaded in various geometric traps. Such
experimental progress allowed a full control of the external parameters in
the Hamiltonian governing the quantum system dynamics. An interesting issue
in the field of ultracold quantum gases is to study the time evolution of a
non-equilibrium situation generated through a quantum quench, which consists
of a sudden change of the Hamiltonian parameters (for example a change of
the harmonic trap frequency or a change in the interaction strength between
the atoms of the gas through Feshbach resonance). A quantum quench is the
easiest way to drive a system to non-equilibrium: the system is supposed to
be in its Hamiltonian ground state until time $t=0$, when the sudden change
of a coupling leads to a new Hamiltonian according to which the system
evolves for $t>0$. On the theoretical side, significant advances have been
carried out in understanding fundamental concepts in the non-equilibrium
dynamics of quantum many-body system. Among these ideas is the link between
quantum dynamics and quantum chaos \cite{Kafri,Borgonovi} and the emergence
of a new ensemble in Statistical Mechanics called generalized Gibbs
ensemble, which is a more general concept than the usual grand-canonical
ensemble and turns out to be a powerful tool in the prediction of relaxation
processes for certain integrable one-dimensional systems 
\cite{Rigol,Rigol2,Caux,Mori,Eckstein,Cardy,Collura}.

In the present work we investigate the quantum dynamics of an ultra-cold
system of $N$ atoms with equal mass $m$ following an interaction quench from
finite to zero interaction strength. We start by writing down the underlying
Hamiltonians before and after the interaction quench. For $t<0$ the gas is
in equilibrium\ and its many-body Hamiltonian is 
\begin{equation}
H_{0}=\sum \limits_{i=1}^{N}\frac{\mathbf{p}_{i}^{2}}{2m}+\sum_{i=1}^{N}V_{0}(\mathbf{r}_{i})+\frac{1}{2}\sum \limits_{i\neq
j}^{N}\sum \limits_{j=1}^{N}v(\mathbf{r}_{i},\mathbf{r}_{j}),  \label{1}
\end{equation}
where $\mathbf{p}_{i}=-i\hbar \mathbf{\nabla }_{i}$ is the momentum operator
of the particle $i$, $V_{0}$ is an external confining potential, and 
$v(\mathbf{r}_{i},\mathbf{r}_{j})$ is a two-body atom-atom interaction. We assume that
the system is initially in a quantum many-body state $\left \vert \Phi
_{0}\right \rangle $ and its associated initial reduced one-body-density
matrix $\rho _{0}(\mathbf{r},\mathbf{r}^{\prime })$ is defined as (see for
instance Ref. \cite{Dreizler}) 
\begin{equation}
\rho _{0}(\mathbf{r},\mathbf{r}^{\prime })=
N\int d\mathbf{r}_{2}d\mathbf{r}_{3}\cdots d\mathbf{r}_{N}\ \Phi _{0}(\mathbf{r},\mathbf{r}_{2},
\mathbf{r}_{3},\dots ,\mathbf{r}_{N})\Phi _{0}^{\ast }(\mathbf{r}^{\prime }, 
\mathbf{r}_{2},\mathbf{r}_{3},\dots ,\mathbf{r}_{N}).  \label{2new}
\end{equation}
At $t=0$, the interactions are turned-off and the many-body Hamiltonian is
given by 
\begin{equation}
H=\sum \limits_{i=1}^{N}\frac{\mathbf{p}_{i}^{2}}{2m}+\sum
\limits_{i=1}^{N}V(\mathbf{r}_{i}),  \label{3new}
\end{equation}%
where we are assuming that the external potential may be different before
and after the interaction quench. At times $t>0$, the system is in a time
dependent quantum state given by 
\begin{equation}
\left \vert \Phi (t)\right \rangle =e^{-\frac{i}{\hbar }Ht}\left \vert \Phi
_{0}\right \rangle .  \label{4new}
\end{equation}%
It should be noted that $\left \vert \Phi _{0}\right \rangle $ is not an
eigenstate of the post-quench Hamiltonian $H$. The latter Hamiltonian
describes a system of independent particles, while $\left \vert \Phi
_{0}\right \rangle $ describes the state of an initially interacting
particles. Before proceeding further, we shall first derive a relationship
between the initial $\rho _{0}(\mathbf{r},\mathbf{r}^{\prime })$, and the
time dependent $\rho (\mathbf{r},\mathbf{r}^{\prime };t)$ reduced
one-body-density matrices. Using the definition 
\begin{equation}
\rho (\mathbf{r},\mathbf{r}^{\prime };t)=N\int 
d\mathbf{r}_{2}d\mathbf{r}_{3}\dots d\mathbf{r}_{N}
\Phi (\mathbf{r},\mathbf{r}_{2},\mathbf{r}_{3},\dots ,\mathbf{r}_{N};t)
\Phi ^{\ast }(\mathbf{r}^{\prime },\mathbf{r}_{2},\mathbf{r}_{3},\dots ,\mathbf{r}_{N};t),  
\label{5new}
\end{equation}
we show in Appendix~A the following relation 
\begin{equation}
\rho (\mathbf{r},\mathbf{r}^{\prime };t)=\int \int U(\mathbf{r},\mathbf{\xi}_{1};t)
\rho _{0}(\mathbf{\xi }_{1},\mathbf{\xi }_{2})\,U^{\ast }(\mathbf{r}
^{\prime },\mathbf{\xi }_{2};t)\ d\mathbf{\xi}_{1}d\mathbf{\xi}_{2},  \label{6new}
\end{equation}%
where $U(\mathbf{r},\mathbf{\xi}_{1};t)=\left \langle \mathbf{r}\right \vert
e^{-\frac{i}{\hbar }(\frac{\mathbf{p}^{2}}{2m}+V)t}\left \vert 
\mathbf{\xi}_{1}\right \rangle $ is the single particle propagator associated to the
post quench Hamiltonian in Eq.~\eqref{3new}. The above relation describes
the time evolution of the one-body density matrix for the considered
specific quench of interaction, when the system is suddenly driven from
interacting to noninteracting configurations.

Let us now come to our main concern, and study the subsequent dynamics of
the system in phase space. For this quench scenario we relate the so-called
dynamical Wigner distribution function to its initial value just before the
quench. Our interest in the Wigner distribution function \cite{Wigner} is
motivated by the fact that it provides a useful tool to study various
properties of many-body systems. Besides, it is well known that it allows a
reformulation of quantum mechanics in terms of classical concepts
\cite{Groenewold,Moyal,Gadella0}, and it is also used to generate semi-classical
approximations \cite{Ozorio,Brack}. Although other distribution functions
exist, the Wigner distribution function has the virtue of its mathematical
simplicity. Nevertheless, it may take negative values as a manifestation of
its quantum nature, and therefore it does not represent a true probability
but a quasi-probability distribution \cite{Hillery,Ozorio}. Wigner
distribution functions have been used in various contexts, as in cold atomic
gases \cite{Giorgini,Bloch}, quantum optics \cite{Walls}, quantum
information \cite{Douce}, quantum chaos \cite{Berry}, and in the study of
non-equilibrium dynamics generated by the perturbation of a Fermi gas system 
\cite{Bettelheim}. Also, the Wigner distribution function of the
noninteracting limit of a Fermi gas system at zero and nonzero temperatures
has been the subject of recent studies \cite{Dean,Zyl,Bencheikh}. The Wigner
function is defined as the Fourier transform of the one-body density matrix $
\rho (\mathbf{r}_{1},\mathbf{r}_{2};t)\equiv \rho (\mathbf{r}+\mathbf{s}/2,
\mathbf{r}-\mathbf{s}/2;t)$ on the relative coordinate $\mathbf{s}=\mathbf{r}
_{1}-\mathbf{r}_{2}$, where $\mathbf{r}=(\mathbf{r}_{1}+\mathbf{r}_{2})/2$
is the centre of mass coordinate. To be precise, we have in $d$ dimensions 
\begin{equation}
W(\mathbf{r},\mathbf{p};t)=\int d\mathbf{s}\, \rho (\mathbf{r}+\mathbf{s}/2,
\mathbf{r}-\mathbf{s}/2;t)\,e^{-i \mathbf{p.s}/\hbar},  \label{8}
\end{equation}%
which is a function of phase space variables $(\mathbf{r},\mathbf{p})$ at
time $t$. The inverse transformation reads 
\begin{equation}
\rho (\mathbf{r}+\mathbf{s}/2,\mathbf{r}-\mathbf{s}/2;t)=\int \frac{d\mathbf{
p}}{(2\pi \hbar )^{d}}\,W(\mathbf{r},\mathbf{p};t)e^{+i \mathbf{p.s}/\hbar}.  \label{9}
\end{equation}%
Setting $\mathbf{s}=\mathbf{0}$ in Eq. \eqref{9}, one recovers the spatial
local density 
\begin{equation}
\rho (\mathbf{r};t):=\rho (\mathbf{r},\mathbf{r};t)=\int \frac{d\mathbf{p}}{
(2\pi \hbar )^{d}}\,W(\mathbf{r},\mathbf{p};t),  \label{10}
\end{equation}%
normalized to the total particle number as $\int \rho (\mathbf{r};t)\,d
\mathbf{r}=N$. Integrating $W(\mathbf{r},\mathbf{p};t)$ over the whole 
space allows to obtain the momentum distribution 
\begin{equation}
n(\mathbf{p};t)=\int \frac{d\mathbf{r}}{(2\pi \hbar )^{d}}\,W(\mathbf{r},%
\mathbf{p};t),\quad \text{such that}\quad \int n(\mathbf{p};t)\,d\mathbf{p}%
=N.  \label{11}
\end{equation}

The structure of the paper is as follows. In Section~\ref{untrapped} we
consider the case of an initially untrapped one dimensional interacting
system subjected to a sudden swich-off of interactions. The resulting
ballistic dynamics is examined in phase space. We derive a relationship
between the dynamical Wigner phase space density at time $t>0$ and its
initial value before the quench. As an application, we analyze the case of a
two-particle system initially interacting through two different zero-range
interactions of Dirac delta type. Section~\ref{oscill} is devoted to a study
of nonequilibrium dynamics through the Wigner function of harmonically
trapped noninteracting particles. Very recently, this system, in one spatial
dimension $(d=1)$, has been the subject of an interesting study \cite{DeanEPL},
where it is suddenly subjected to a modification of the trapping frequency and a relationship
between the resulting time dependent Wigner function and its initial value
is obtained. In this section we propose a generalization to arbitrary
spatial dimensions $(d\geq 1)$ of this relation by using an alternative
fully quantum mechanical method which is based on the use of time evolution
of the one-body density matrix. As a bonus, by using the inverse Wigner
transformation to our generalized relationship in Wigner phase space, we
derive the scaling law of harmonically trapped noninteracting particles in
arbitrary dimension $d$, between the one-body density matrix at time $t$ and
its initial value. For $d=1$, our result reduces to the one obtained in \cite{DeanEPL}, as it should.

In order to observe how the dynamical Wigner function following a quench is
affected by interactions, we consider physical systems of
particles whose dynamics are governed by scaling laws suffering a
non-ballistic expansion, and we compare the resulting Wigner function with
the one obtained for a ballistic expansion. 
Some final conclusions put an end to the paper in Section~\ref{remarks}.

\section{Time evolution of Wigner function of an initially untrapped
interacting system following a sudden swich-off of interactions.}
\label{untrapped}

In this section let us consider the situation where both external potentials $V_{0}$ and $V$
 in Eqs.~\eqref{1} and \eqref{3new} vanish,
and the study is restricted to one-dimensional interacting particles system.
The time evolution of the resulting reduced one-body density matrix after a
finite time of free expansion $t>0$ is then given by the one dimensional
version of Eq. \eqref{6new}, that is 
\begin{equation}
\rho (x_{1},x_{2};t)=\int_{-\infty }^{\infty }\int_{-\infty }^{\infty
}U(x_{1},\xi _{1};t)\, \rho _{0}(\xi _{1},\xi _{2})\,U^{\ast }(x_{2},\xi
_{2};t)\ d\xi _{1}d\xi _{2}.  \label{3}
\end{equation}%
Here $U(x,\xi ;t)=\left \langle x\right \vert e^{-itH_{0}/\hbar }\left \vert
\xi \right \rangle $ with $H_{0}=-\frac{\hbar ^{2}}{2m}\frac{d^{2}}{dx^{2}}$%
. The matrix element $U(x,\xi ;t)$ is the free particle Feynman propagator
in the configuration space, given by \cite{Feynman} 
\begin{equation}
U(x,\xi ;t)=\sqrt{\frac{m}{2\pi i\hbar t}}\ \exp \left( i\frac{m}{2\hbar t}%
(x-\xi )^{2}\right) .  \label{4}
\end{equation}%
If we substitute this expression into Eq. \eqref{3} we find 
\begin{equation}
\rho (x_{1},x_{2};t)=\frac{m}{2\pi \hbar t}\int_{-\infty }^{\infty
}\int_{-\infty }^{\infty }e^{i\frac{m}{2\hbar t}\left[ (x_{1}-\xi
_{1})^{2}-(x_{2}-\xi _{2})^{2}\right] }\rho _{0}(\xi _{1},\xi _{2})\,d\xi
_{1}d\xi _{2},  \label{5}
\end{equation}%
which can be rewritten as 
\begin{equation}
\rho (x_{1},x_{2};t)=\frac{m}{2\pi \hbar t}e^{i\frac{m}{2\hbar t}%
(x_{1}^{2}-x_{2}^{2})}\int_{-\infty }^{\infty }\int_{-\infty }^{\infty }e^{i%
\frac{m}{2\hbar t}\left[ (\xi _{1}-\xi _{2})(\xi _{1}+\xi _{2})-2x_{1}\xi
_{1}+2x_{2}\xi _{2})\right] }\rho _{0}(\xi _{1},\xi _{2})\ d\xi _{1}d\xi
_{2}.  \label{6}
\end{equation}%
Now it is more convenient to introduce the center-of-mass and relative
coordinates, respectively defined by $z=(\xi _{1}+\xi _{2})/2$ and $%
z^{\prime }=\xi _{1}-\xi _{2}$. Then we can write 
\begin{equation}
\rho (x_{1},x_{2};t)=\frac{m\,e^{i\frac{m}{2\hbar t}(x_{1}^{2}-x_{2}^{2})}}{%
2\pi \hbar t}\int_{-\infty }^{\infty }\int_{-\infty }^{\infty }e^{i\frac{m}{%
\hbar t}zz^{\prime }}\,e^{-i\frac{m}{\hbar t}(x_{1}-x_{2})z}\,e^{-i\frac{m}{%
2\hbar t}(x_{1}+x_{2})z^{\prime }}\, \rho _{0}(z+\tfrac{z^{\prime }}{2},z-%
\tfrac{z^{\prime }}{2})\,dzdz^{\prime }.  \label{7}
\end{equation}%
In the following we shall examine the time evolution of the above one-body
density matrix in the Wigner representation.

To proceed with computing the Wigner distribution function, we insert Eq. %
\eqref{7} into \eqref{8}, and we may write 
\begin{align*}
W(x,p;t)& =\frac{m}{2\pi \hbar t}\int_{-\infty }^{\infty }ds\,e^{i\frac{m}{%
\hbar t}xs}\int_{-\infty }^{\infty }\int_{-\infty }^{\infty }e^{i\frac{m}{%
\hbar t}zz^{\prime }}\,e^{-i\frac{m}{\hbar t}zs}\,e^{-i\frac{m}{\hbar t}%
z^{\prime }x}\,e^{-ips/\hbar }\rho _{0}(z+\tfrac{z^{\prime }}{2},z-\tfrac{%
z^{\prime }}{2})\,dzdz^{\prime } \\
& =\frac{m}{2\pi \hbar t}\int_{-\infty }^{\infty }\int_{-\infty }^{\infty
}e^{i\frac{m}{\hbar t}zz^{\prime }}\,e^{-i\frac{m}{\hbar t}z^{\prime }x}\,
\rho _{0}(z+\tfrac{z^{\prime }}{2},z-\tfrac{z^{\prime }}{2})\,dzdz^{\prime
}\int_{-\infty }^{\infty }ds\,e^{\frac{i}{\hbar }\left[ \frac{m}{t}x-p-\frac{%
m}{t}z\right] s}.
\end{align*}%
Carrying out the integration on the variable $s$, we obtain 
\begin{equation}
W(x,p;t)=\int_{-\infty }^{\infty }\int_{-\infty }^{\infty }e^{i\frac{m}{%
\hbar t}zz^{\prime }}e^{-i\frac{m}{\hbar t}xz^{\prime }}\, \delta \left( x-%
\tfrac{pt}{m}-z\right) \, \rho _{0}(z+\tfrac{z^{\prime }}{2},z-\tfrac{%
z^{\prime }}{2})\,dzdz^{\prime },  \label{13}
\end{equation}%
which after integration over $z$ reduces to 
\begin{equation}
W(x,p;t)=\int_{-\infty }^{\infty }\rho _{0}(x-\tfrac{pt}{m}+\tfrac{z^{\prime
}}{2},x-\tfrac{pt}{m}-\tfrac{z^{\prime }}{2})\,e^{-ipz^{\prime }/\hbar
}\,dz^{\prime }.  \label{14}
\end{equation}%
But according to Eq. \eqref{8}, the above integral is nothing but the
initial Wigner distribution function at $t=0$, and at the phase space point $%
\left( x-pt/m,p\right) $, so that for $t>0$ 
\begin{equation}
W(x,p;t)=W_{0}(x-\tfrac{pt}{m},p).  \label{15}
\end{equation}%
This relation is the Wigner phase space version of Eq. \eqref{3}. Upon
integration over $x$ and making use of the definition in Eq. \eqref{11}, we
get for $t>0$ 
\begin{equation}
n(p,t)=\int_{-\infty }^{\infty }\frac{dx}{2\pi \hbar }\,W(x,p;t)=\int_{-%
\infty }^{\infty }\frac{dx}{2\pi \hbar }W_{0}(x-\tfrac{pt}{m}%
,p)=\int_{-\infty }^{\infty }\frac{d\xi }{2\pi \hbar }W_{0}(\xi ,p)=n(p,0).
\label{16}
\end{equation}%
Therefore, we find that under ballistic expansion (where interatomic
collisions during the expansion are not present) the dynamical momentum
density for $t>0$ is equal to its initial value $n(p,0)$. Some information
can be obtained from this result. For example, if one considers an initial
quantum gas with short range interaction, whose momentum distribution $%
n(p,0) $ exhibits a $1/p^{4}$ tail at large momentum (see for instance Ref.~%
\cite{Olshanii}), equation \eqref{16} says that this long tail behavior is
preserved in the dynamical momentum density for all positive times.

It is worth noticing that Eq. \eqref{3} remains valid for an initial
non-vanishing confining potential $(V_{0}\neq 0)$ provided that one realizes
at $t=0$ a simultaneously sudden double quench, where the interactions are
turned-off with the release of the trap $(V=0)$. As a consequence the
results in Eqs.~\eqref{15} and \eqref{16}, in this case hold also true.

\subsection{Case of two particles interacting through an attractive $\protect%
\delta$ interaction}

\begin{figure}[htbp]
\centering
\includegraphics[width=7cm]{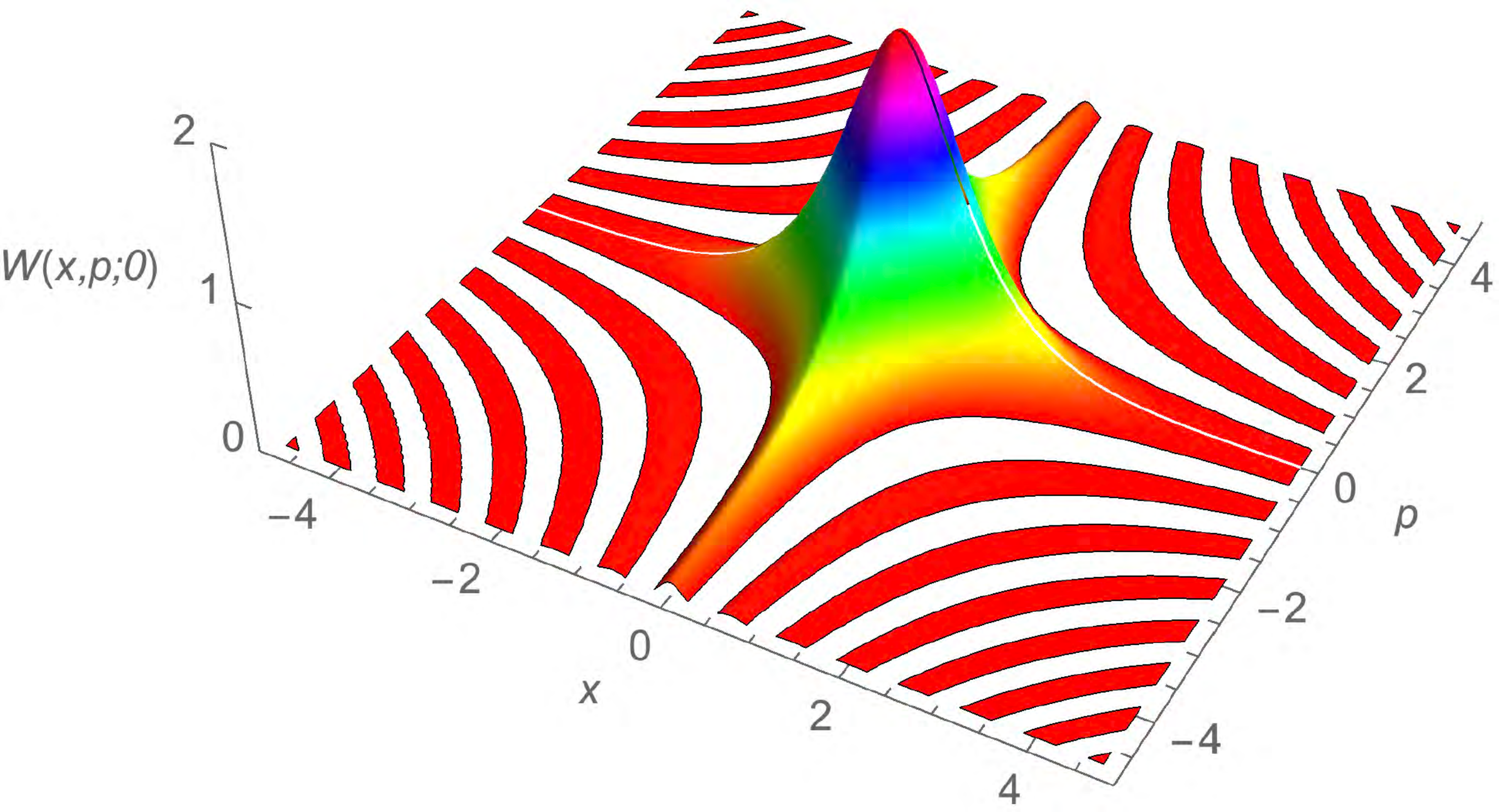} \qquad\qquad %
\includegraphics[width=7cm]{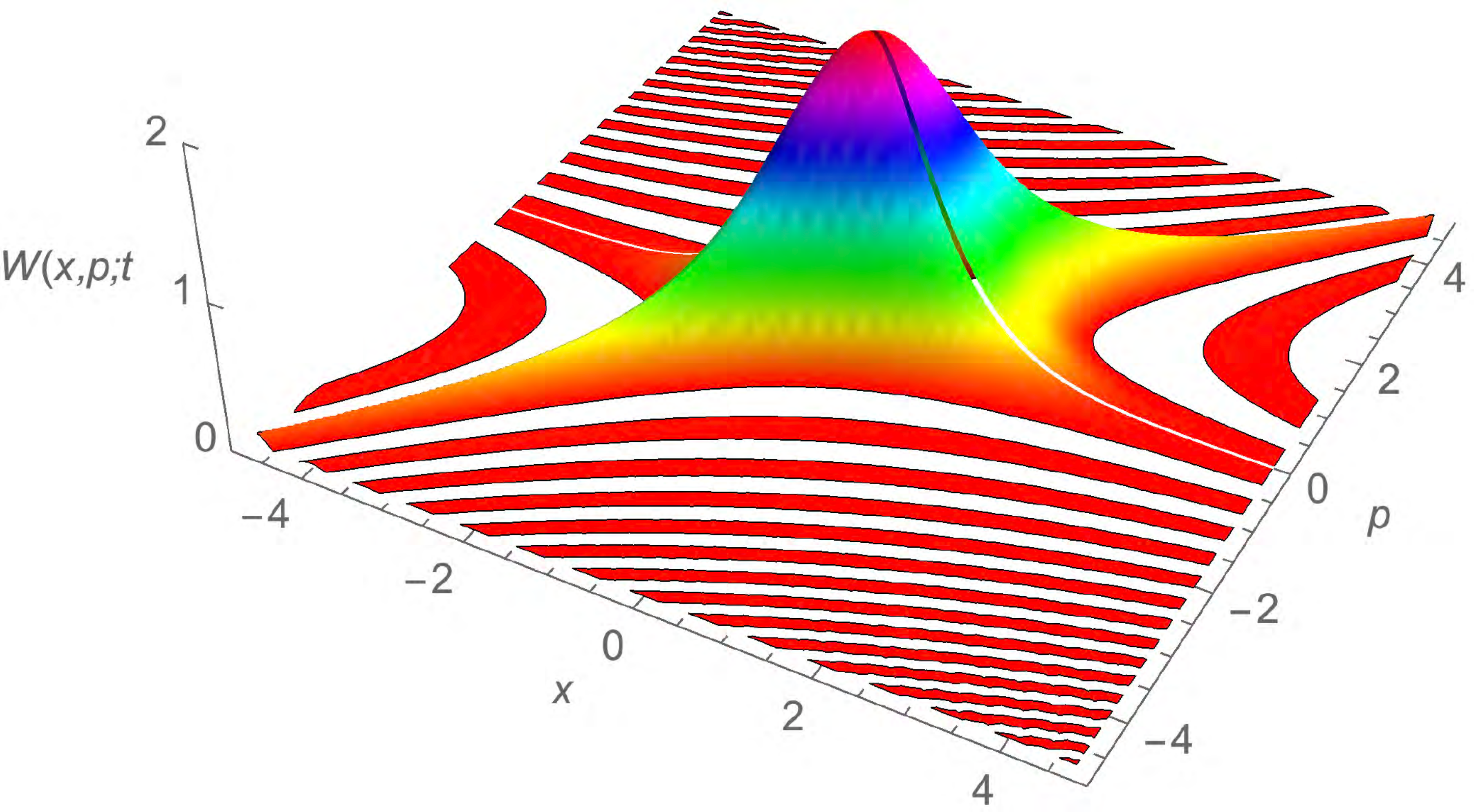}
\caption{{\protect\small The Wigner function for an attractive Dirac $%
\protect\delta$ interaction obtained in Eq.~\eqref{20}, for the values $%
\hbar=m=a=1$. On the left for $t=0$, where two symmetry axes, $x=0$ and $p=0$%
, are evident; on the right after the quenching, for $t=1$, a clear
distortion of the initial function can be appreciated: the perfect symmetry
is lost and squeezing appears between the axes $p=0$ and $p=x$. The white
regions correspond to small negative values of $W(x,p;t)$.}}
\label{fig_1}
\end{figure}

As a first case study we consider a simple model system of two particles
interacting through an attractive Dirac delta potential with highly
asymmetric mass imbalance: the particle with mass $M$ is so heavy that the
center-of-mass motion can be ignored. The problem reduces then to a one-body
system and the $m$ mass light particle Hamiltonian is given for $t\leq 0$ by 
\begin{equation}
H_{a}=-\frac{\hbar ^{2}}{2m}\frac{d^{2}}{dx^{2}}-a\delta (x).  \label{18}
\end{equation}%
Here $a>0$ is the strength of the interaction. The single normalized bound
state is $\phi (x)=\sqrt{\alpha }\,e^{-\alpha \left \vert x\right \vert }$
with energy $E_{a}=-ma^{2}/(2\hbar ^{2})$, and $\alpha =ma/\hbar ^{2}$. The
corresponding Wigner distribution function for $t\leq 0$ can be calculated
analytically using \eqref{8} and is found to be 
\begin{equation}
W_{0}(x,p)=\frac{2\alpha ^{2}\hbar ^{2}\,e^{-2\alpha \left \vert x\right
\vert }}{p^{2}+\alpha ^{2}\hbar ^{2}}\left[ \cos \left( \frac{2p\left \vert
x\right \vert }{\hbar }\right) +\frac{\alpha \hbar }{p}\sin \left( \frac{%
2p\left \vert x\right \vert }{\hbar }\right) \right] .  \label{19}
\end{equation}%
A plot of this function is given on the left hand side of Figure~\ref{fig_1}.

At time $t=0$ the interaction is suddenly turned off ($a=0$) and then, from %
\eqref{15} we get the following Wigner function for $t>0$: 
\begin{equation}
W(x,p;t)=\frac{2\alpha^{2}\hbar^{2}\, e^{-2\alpha \left \vert x-\frac{p}{m}%
t\right \vert }}{p^{2}+\alpha^{2}\hbar^{2}}\left[ \cos \left( \frac{2p\left
\vert x-\frac{p}{m}t\right \vert }{\hbar}\right) +\frac{\alpha \hbar}{p}\sin
\left( \frac{2p\left \vert x-\frac{p}{m}t\right \vert }{\hbar}\right) \right]%
.  \label{20}
\end{equation}
A plot of this function $W(x,p;t)$ is given for a particular value of $t>0$
on the right hand side of Figure~\ref{fig_1}. A clear distortion can be
observed as time grows: for $t=0$ there are two symmetry axes, $x=0$ and $%
p=0 $; for bigger values of $t$ the perfect symmetry is lost, although the
axes $p=0$ and $p=mx/t$ (this one clockwise rotating with time) play an
important role; squeezing is more and more pronounced as $t\to \infty$, the
axis $p=mx/t $ approaching the axis $p=0$.

Finally, using \eqref{16} it is easy to show analytically that, the initial
momentum density is 
\begin{equation}  \label{np0delta}
n(p,0)=\frac{2 \alpha^3 \hbar^3}{\pi(p^2+\alpha^2 \hbar^2)^2},
\end{equation}
It is easy to check that this result satisfies the second equation in \eqref{11} for $N=1$ particle.
It has been shown in \cite{Olshanii} that for a system of zero-range $\delta
-$interacting one-dimensional atoms with arbitrary strength, the high-$p$
asymptotic behavior of the momentum distribution for both free and
harmonically trapped atoms, exhibits a universal $1/p^{4}$ dependence. As
can be seen in Eq. \eqref{np0delta} and at large values of the momentum $p$,
we recovered this $1/p^{4}$ dependence of the momentum distribution.

\subsection{Case of two particles interacting through a $\protect\delta$-$%
\protect\delta^{\prime }$ interaction}

\begin{figure}[htbp]
\centering
\includegraphics[width=7cm]{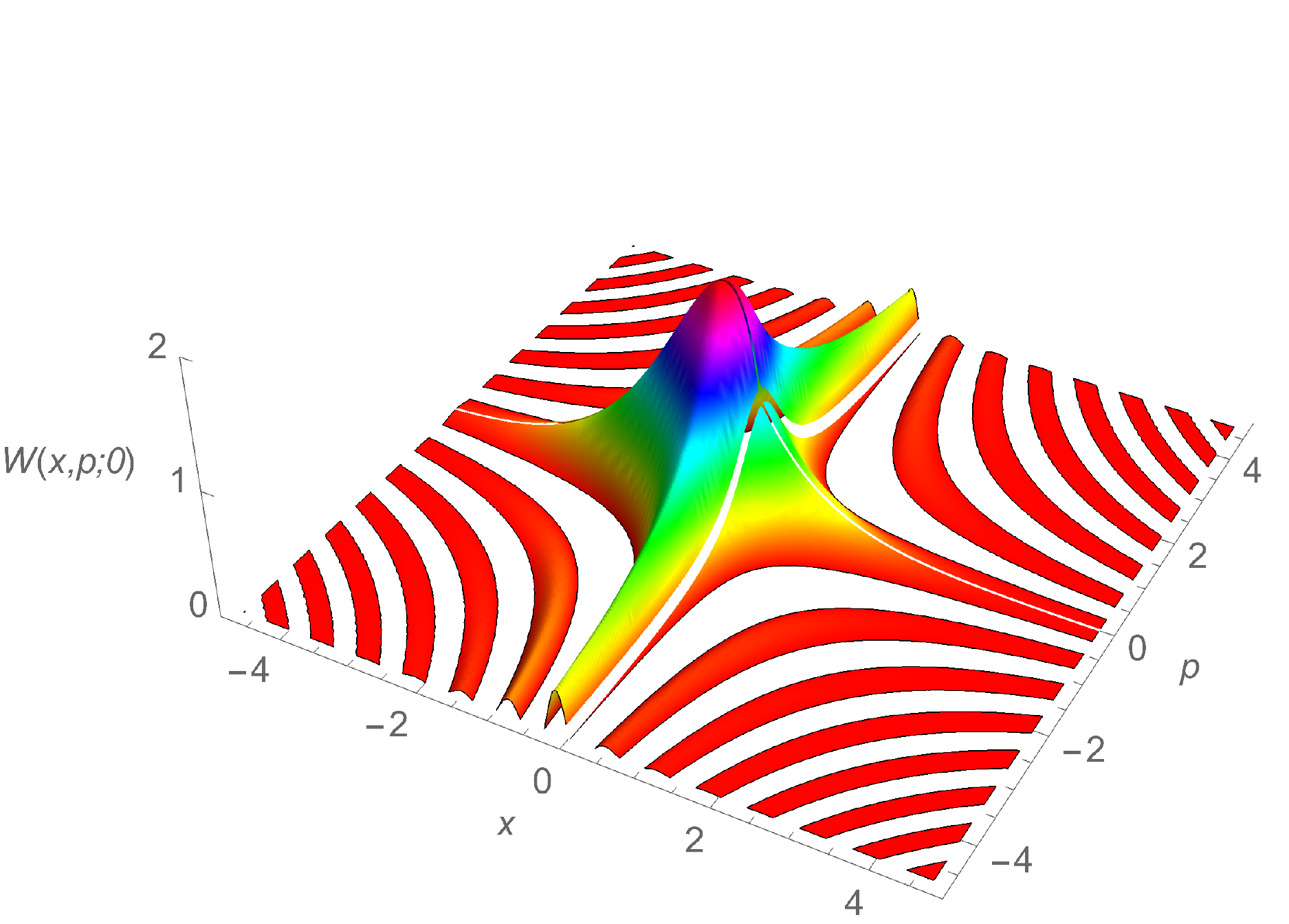} \qquad\qquad %
\includegraphics[width=7cm]{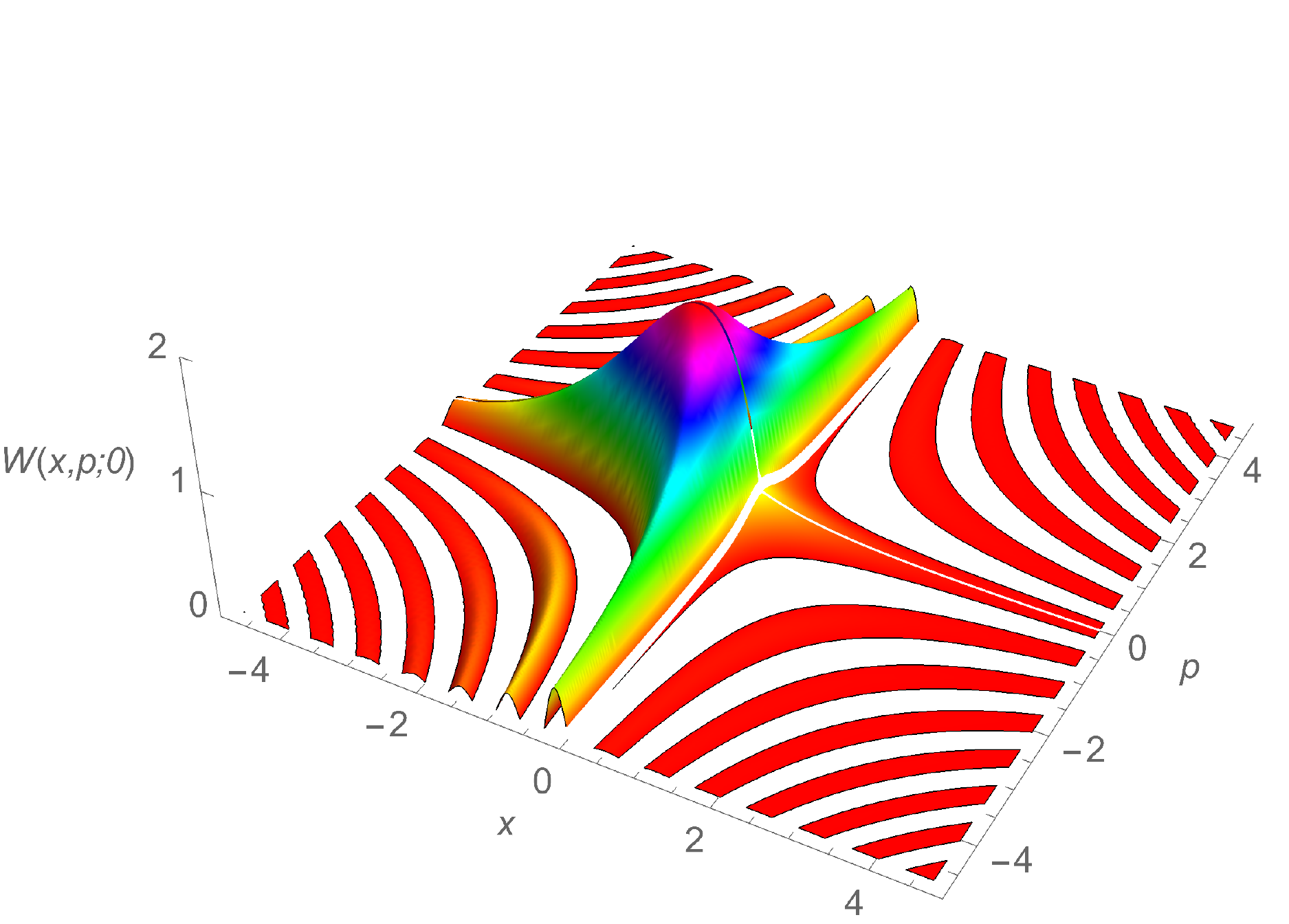} \\[4ex]
\includegraphics[width=7cm]{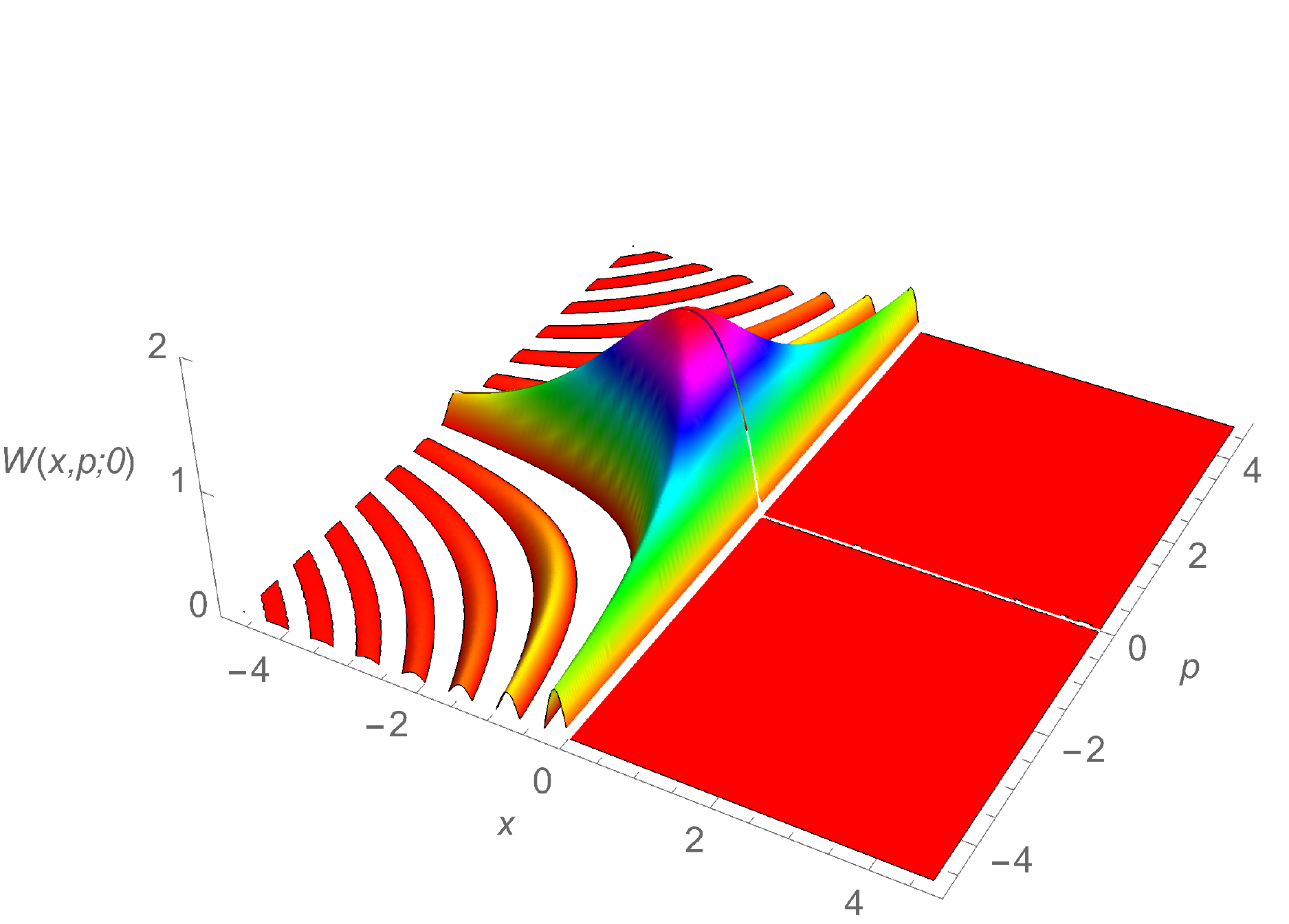} \qquad\qquad %
\includegraphics[width=7cm]{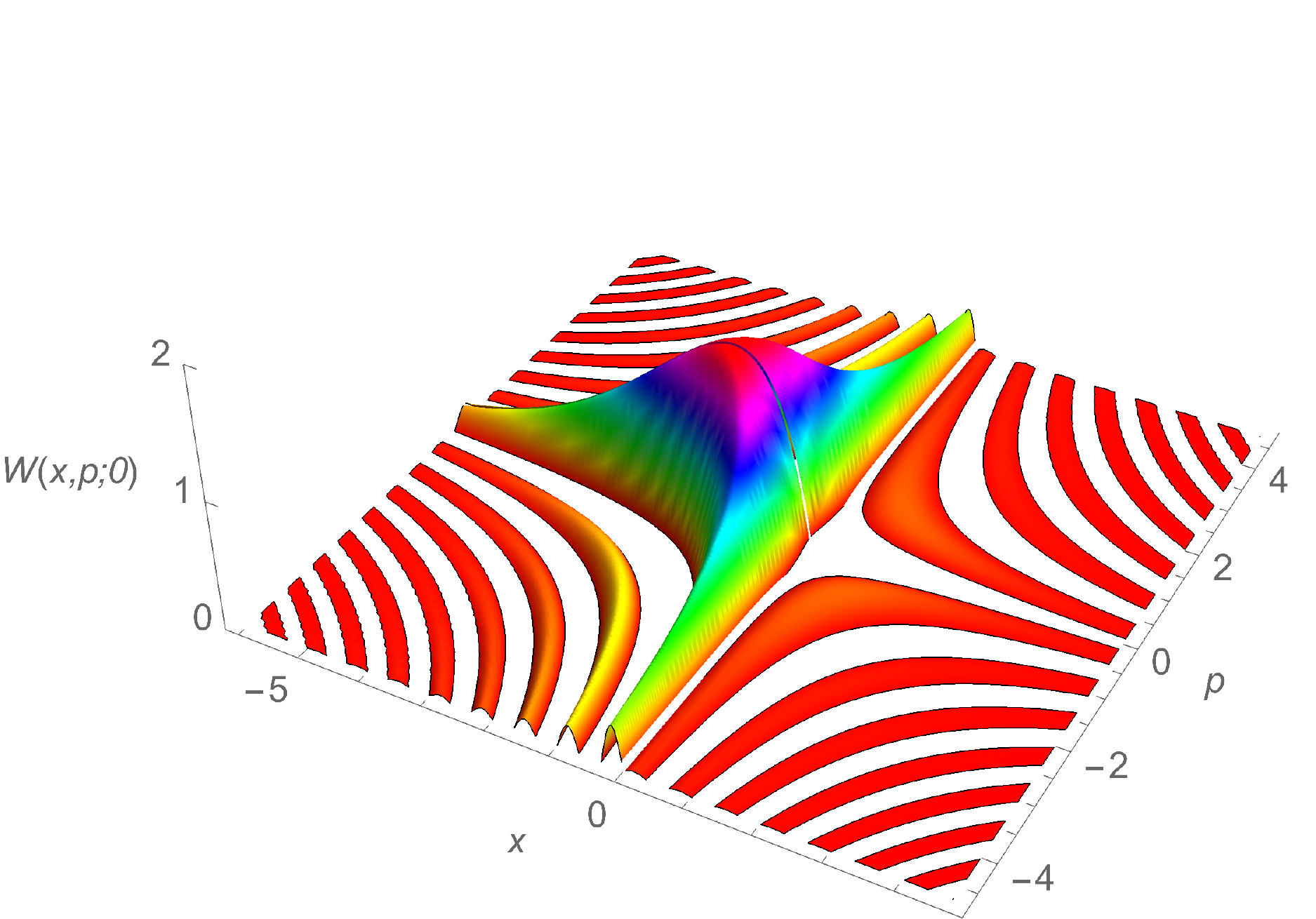}
\caption{{\protect\small The Wigner function for a Dirac $\protect\delta$-$%
\protect\delta^{\prime }$ interaction as in \eqref{b1}, for the values $%
\hbar=m=a=1$. From left to right, and from top to bottom: $b=-0.6$, $b=-0.9$%
, $b=-1.0$ and $b=-1.2$. The white regions correspond to small negative
values of $W_0(x,p)$ in \eqref{b5}. The discontinuity of the surface at $x=0$ can be
clearly seen on the plots.}}
\label{fig_2}
\end{figure}

We are going to consider now an extension of the previous study of two
interacting particles that takes into account the presence of an extra
point-like interaction term in the potential, proportional to $\delta
^{\prime }$. This type of point or zero-range potentials are a subject of
recent study in differents contexts \cite%
{Gadella1,Gadella2,Gadella3,Romaniega}. The Hamiltonian of the light
particle with mass $m$ is now given for $t<0$ by 
\begin{equation}
H_{a,b}=-\frac{\hbar ^{2}}{2m}\frac{d^{2}}{dx^{2}}-a\delta (x)+b\delta
^{\prime }(x),\quad a>0,\ b\in \mathbb{R}.  \label{b1}
\end{equation}%
The associated Schr\"{o}dinger equation has been carefully analyzed in \cite%
{Gadella1}, where it was proven that the above Hamiltonian supports only one
bound state of energy 
\begin{equation}
E_{a,b}=-\frac{ma^{2}}{2\hbar ^{2}\left( 1+\frac{m^{2}b^{2}}{\hbar ^{4}}%
\right) ^{2}}=-\frac{\hbar ^{2}\, \kappa ^{2}}{2m},\quad \kappa >0.
\label{b2}
\end{equation}%
The normalized wave function is 
\begin{equation}
\phi _{a,b}(x)=A\,e^{-\kappa |x|}\left( 1+B\, \text{sign}(x)\right) ,
\label{b3}
\end{equation}%
where $\text{sign}(x)$ stands for the sign function and
\begin{equation}
A=\frac{\sqrt{ma}}{\hbar (1+B^{2})},\qquad B=\frac{mb}{\hbar ^{2}}.
\label{b4}
\end{equation}%
In this case
the Wigner distribution function for $t\leq 0$ can be also determined
analytically from \eqref{8} and it turns out to be the following expression 
\begin{eqnarray}\label{b5} 
W_{0}(x,p)=\frac{2\kappa^2\hbar ^{2}\,e^{-2\kappa |x|}}{%
(1+B^2)(p^{2}+\kappa ^{2}\hbar ^{2})} 
&&
\left[ (1-B^{2}) \cos \left( \frac{2p\left
\vert x\right \vert }{\hbar }\right)  \right. \nonumber \\
&& 
\qquad 
\left. +\frac{  \kappa ^{2}\hbar ^{2}+ B^2(2 p^{2}+\kappa ^{2}\hbar ^{2}) +2B(p^{2}+\kappa ^{2}\hbar ^{2})\, \text{sign}(x) }
{\kappa\hbar p}
\sin \left( \frac{2p|x|}{\hbar }\right) %
\right] .  
\end{eqnarray}%
Remark that the presence of the sign function on \eqref{b5} indicates the
presence of a discontinuity of the Wigner function $W(x,p;0)$ along the line 
$x=0$. If we consider the limit $B\to 0$ in the last expression we recover the result of equation \eqref{19}, as one should expect.
Some plots of this function are given in Figure~\ref{fig_2} for $%
b=-0.6$, $b=-0.9$, $b=-1.0$ and $b=-1.2$.

\begin{figure}[htbp]
\centering
\includegraphics[width=7cm]{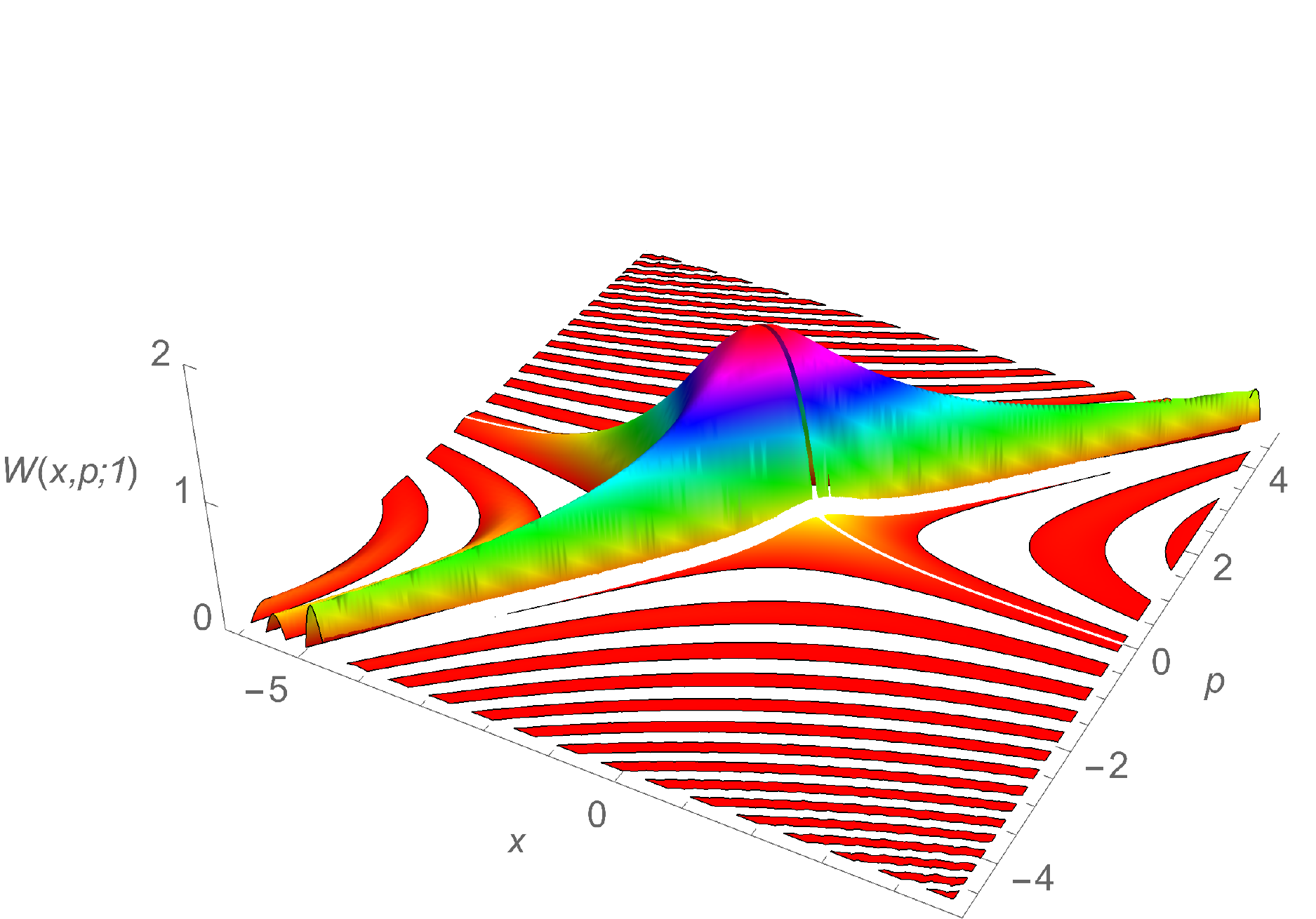} \qquad\qquad %
\includegraphics[width=7cm]{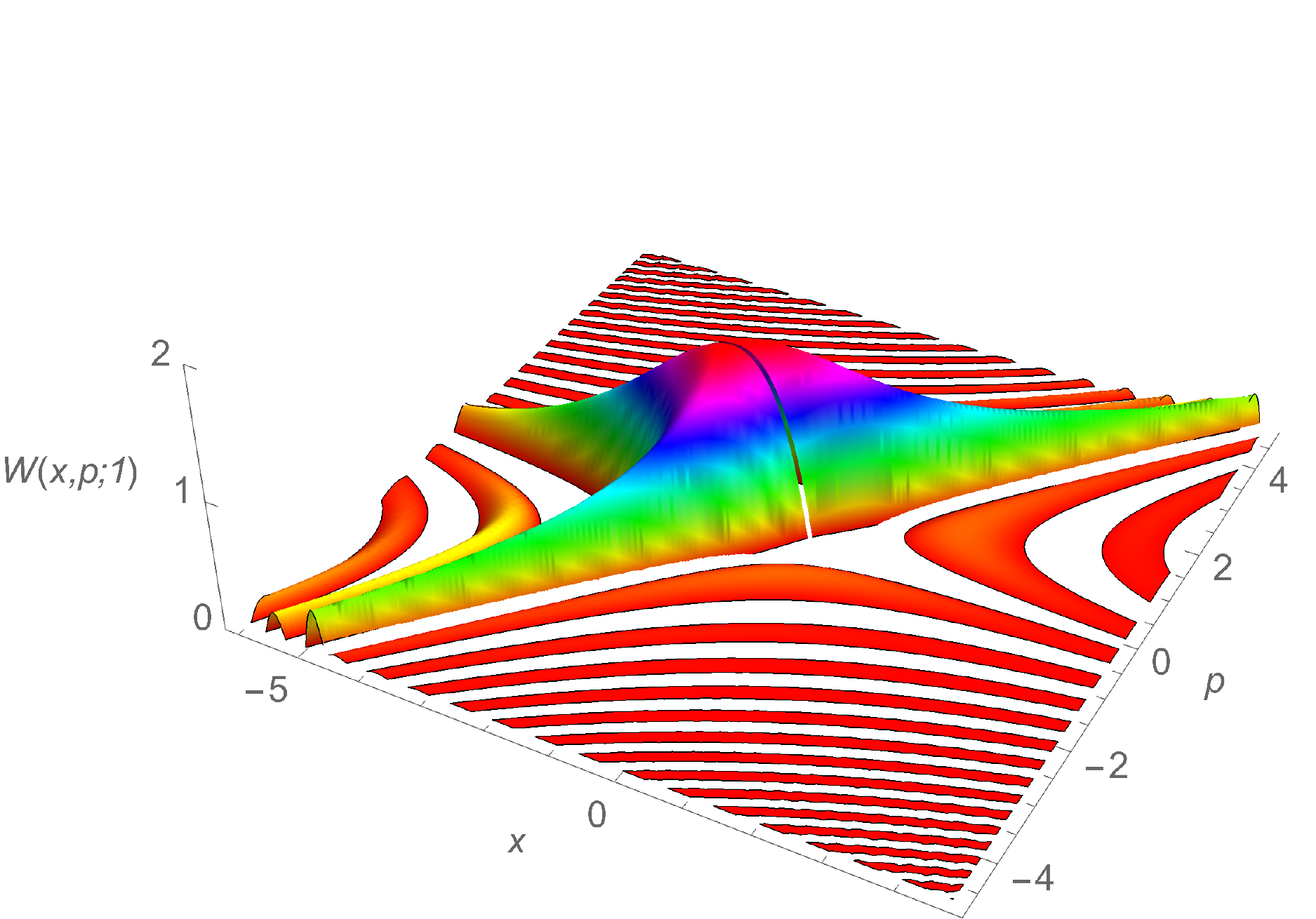}
\caption{{\protect\small The Wigner function $W(x,p;t=1)$ for a Dirac $%
\protect\delta$-$\protect\delta^{\prime }$ interaction with values $%
\hbar=m=a=1$, for $b=-0.9$ (left) and $b=-1.2$ (right), as given by \eqref{b6}. Squeezing of the
initial function Figure~\protect\ref{fig_2} can be appreciated between the
axes $p=0$ and $p=x$. The white regions correspond to small negative values
of $W(x,p;t=1)$. The discontinuity of the surface at $x=p$ can be clearly
seen.}}
\label{fig_3}
\end{figure}

At time $t=0$ the interaction is turned off ($a=b=0$) and then, from %
\eqref{15} we get the following Wigner function for $t>0$: 
 \begin{eqnarray}  \label{b6}
 W(x,p;t)=\frac{2\kappa^2\hbar ^{2}\,e^{-2\kappa \left| x-\tfrac{pt}{m}\right| }}{%
(1+B^2)(p^{2}+\kappa ^{2}\hbar ^{2})} 
&&
\left[ (1-B^{2}) \cos \left( \frac{2p\left
\vert x-\tfrac{pt}{m}\right \vert }{\hbar }\right)  \right. \\
&& 
\qquad 
\left. +\frac{  \kappa ^{2}\hbar ^{2}+ B^2(2 p^{2}+\kappa ^{2}\hbar ^{2}) +2B(p^{2}+\kappa ^{2}\hbar ^{2})\, \text{sign}(x-\tfrac{pt}{m}) }
{\kappa\hbar p}
\sin \left( \frac{2p|x-\tfrac{pt}{m}|}{\hbar }\right) %
\right] . \notag
\end{eqnarray}
Plots of this function $W(x,p;t)$ are given in Figure~\ref{fig_3} for $t=1$
and values $b=-0.9$ (left) and $b=-1.2$ (right). A clear distortion can be
observed comparing with Figure~\ref{fig_2}: the axis $p=0$ is preserved, but
the axis $x=0$ rotates around the origin and becomes $x=p$ (this one
changing with time); squeezing is stronger as $t\to \infty$, the axis $x=0$
approaching the axis $p=0$.

\begin{figure}[htbp]
\centering
\includegraphics[width=7cm]{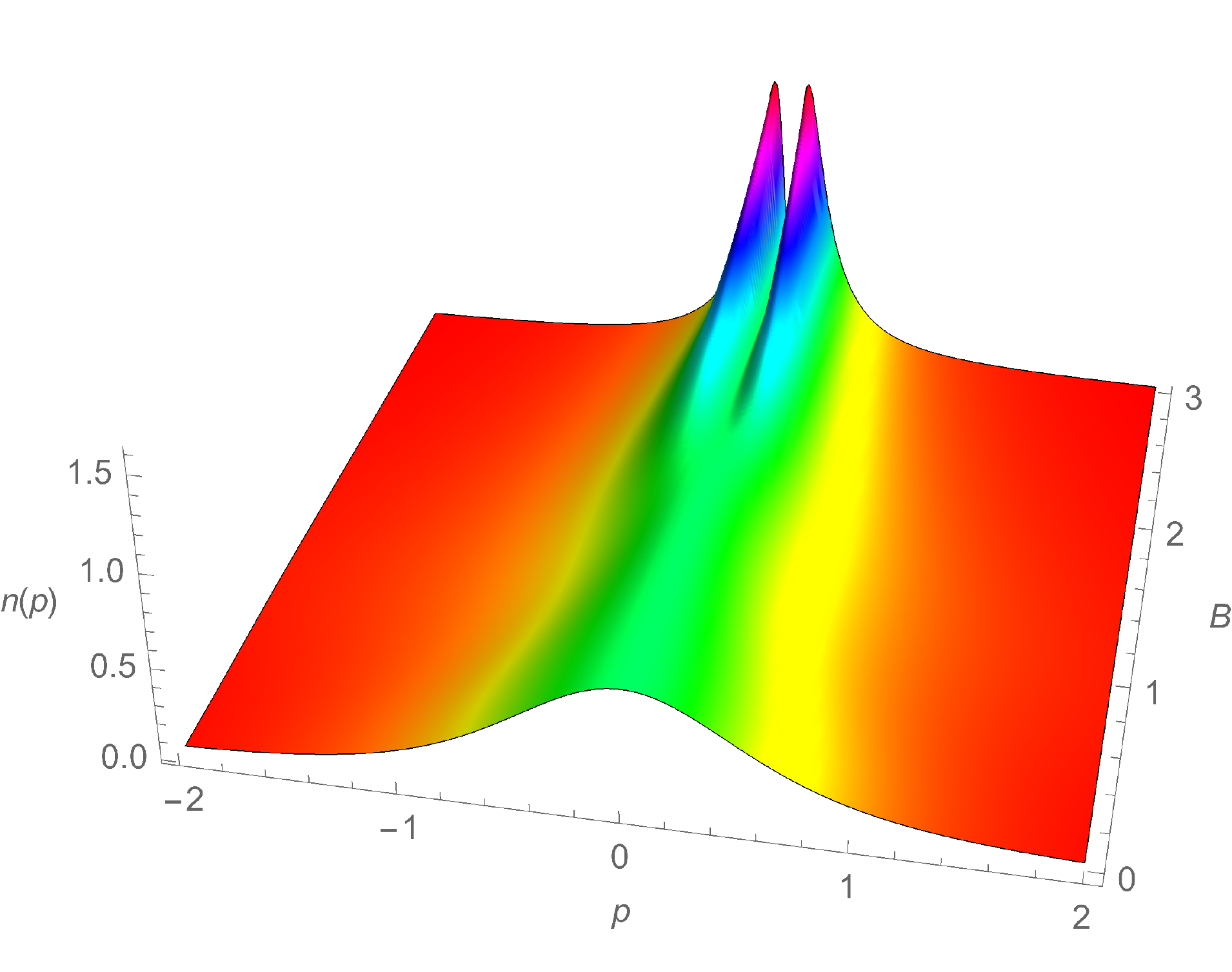}
\caption{{\protect\small Momentum density $n(p)$ as given by \eqref{np0deltaprime} as a function of $p$ and $B\geq 0$, with values $\hbar=m=a=1$. 
For $B=0$ we recover the profile of \eqref{np0delta}, which is similar for $B^2\leq 2$. For values such that $B^2>2$ a peculiar structure appears around $p=0$: $p=0$ turns from a maximum to a minimum and two symmetrical maxima appear at the points $p=\pm \hbar\alpha \sqrt{B^2 - 2}/(B (1 + B^2))$.}}
\label{fig_4}
\end{figure}

From \eqref{16} and \eqref{b5} the initial momentum density can be
determined analytically: 
\begin{equation}  \label{np0deltaprime}
n(p,0)=\frac{2\hbar \alpha}\pi \ 
\frac{B^2 (1 + B^2)^2 p^2 +\hbar^2\alpha^2 }{((1+ B^2)^2 p^2+\hbar^2\alpha^2 )^2}.
\end{equation}
Again, it is easy to check that this result satisfies the second equation in \eqref{11} for $N=1$ particle.
In addition it coincides with \eqref{np0delta} in the limit $B=mb/\hbar^2\to0$ and
which scales as $1/p^2$ for high $p$ if $B\neq 0$ (if $B=0$ it was already
mentioned after \eqref{np0delta} that it scales as $1/p^4$). 
Hence, it is
interesting to observe that the presence of the additional $\delta^{\prime }$
interaction term, changes substantially the high momentum tail of the
momentum density: at large values of the momentum $p$, the momentum
distribution in Eq. \eqref{np0deltaprime} exhibits $1/p^{2}$ behavior while
in the absence of $\delta^{\prime }$ interaction this density scales as $1/p^{4}$.
A plot of the momentum density as a function of $p$ and $B$ in Figure~\ref{fig_4} clearly
shows two effects. The aforementioned behavior at large values of $p$ and the
appearance of a local minimum around $p = 0$. This minimum is due to the presence
of a polynomial in the numerator of the expression of $n(p)$ generated by the
additional $\delta'$ interaction term. 
We observe that this minimum is more pronounced as $B$ increases and transforms to a maximum for smaller $B$ values. 
To be more precise, the point $p=0$ is a minimum of $n(p)$ for $B^2> 2$ and a maximum otherwise. In addition, if $B^2> 2$ two symmetrical maxima appear at the points 
$$
p=\pm \frac{\hbar\alpha \sqrt{B^2 - 2}}{B (1 + B^2)}.
$$

\section{Time evolution of Wigner distribution function following a sudden
change of trapping harmonic potential in d dimensions}
\label{oscill}

Very recently Dean et al. \cite{DeanEPL} 
computed the time evolution of the
Wigner function $W(x,p;t)$ for a one-dimensional system of noninteracting
fermions (called also system of independent fermions), initially trapped by
a harmonic oscillator potential with frequency $\omega _{0}$, subjected to a
sudden change of frequency passing from $\omega _{0}$ to $\omega $. At time $%
t$, after this quench it was found (see Eq. (13) of \cite{DeanEPL}) that
\begin{equation}\label{yy}
W(x,p;t)=W_{0}\left( x\cos \omega t-\frac{p\sin \omega t}{m\omega },m\omega
x\sin \omega t+p\cos \omega t\right) ,
\end{equation}%
where $W_{0}$ is the initial Wigner function of the system. To derive this relationship
 the Liouville equation is used. Here we propose an alternative fully quantum mechanical derivation
based on the use of time evolution of one-body density matrix which is valid
for an arbitrary spatial dimension $d$, which obviously reduces to the
result in \eqref{yy} for $d=1$ as it should. 

Consider a system of noninteracting fermions initially confined in a $d-$%
dimensional isotropic harmonic oscillator potential of the form $V_{0}(%
\mathbf{r})=m\omega _{0}^{2}\mathbf{r}^{2}/2$. Suddenly the frequency is
changed from $\omega _{0}$ to $\omega $ and the system expands in the
potential $V(\mathbf{r})=m\omega ^{2}\mathbf{r}^{2}/2$. According to \eqref{6new}, the time evolution of the resulting one-body density matrix at $t>0$ is
then given by 
\begin{equation}
\rho (\mathbf{r}_{1},\mathbf{r}_{2};t)=\int \int U_{osc}(\mathbf{r}_{1},%
\mathbf{\xi }_{1};t)\rho _{0}(\mathbf{\xi }_{1},\mathbf{\xi }_{2})\,
U_{osc}^{\ast }(\mathbf{r}_{2},\mathbf{\xi }_{2};t)\ d\mathbf{\xi }_{1}d\mathbf{\xi }_{2},
\end{equation}%
Here $\rho _{0}(\mathbf{\xi }_{1},\mathbf{\xi }_{2})$ is the one-body
density matrix at $t=0$ \ and $U_{osc}(\mathbf{r}_{1},\mathbf{\xi }_{1};t)$
is the well known propagator associated to the isotropic harmonic oscillator
potential with frequency $\omega $, given in $d$ dimensions by \cite{Feynman}
\begin{equation}
U_{osc}(\mathbf{r}_{1},\mathbf{\xi }_{1};t)=\left( \frac{m\omega }{2\pi
i\hbar \sin \omega t}\right) ^{\frac{d}{2}}\exp \left[ i\frac{m\omega }{%
2\hbar \sin \omega t}[(\mathbf{r}_{1}^{2}+\mathbf{\xi }_{1}^{2})\cos \omega
t-2\mathbf{r}_{1}\mathbf{ \cdot \mathbf{\xi }_{1}}]\right]  \label{22}
\end{equation}%
by substitution we obtain 
\begin{equation*}
\rho (\mathbf{r}_{1},\mathbf{r}_{2};t)=\left( \frac{m\omega }{2\pi \hbar
\left \vert \sin \omega t\right \vert }\right) ^{d}\int \int \rho _{0}(%
\mathbf{\xi }_{1},\mathbf{\xi }_{2})e^{\frac{im\omega }{2\hbar \sin \omega t}%
[(\mathbf{r}_{1}^{2}+\mathbf{\xi }_{1}^{2}-\mathbf{r}_{2}^{2}-\mathbf{\xi }%
_{2}^{2})\cos \omega t-2\mathbf{r}_{1}\mathbf{.\mathbf{\xi }_{1}}+2\mathbf{r}%
_{2} \cdot \mathbf{\mathbf{\xi }_{2}}]}d\mathbf{\xi }_{1}d\mathbf{\xi }_{2}.
\end{equation*}%
Using the centre of mass $\mathbf{u}=(\mathbf{\xi }_{1}+\mathbf{\xi}_{2})/2$ and relative
$\mathbf{v}=(\mathbf{\xi }_{1}-\mathbf{\xi }_{2})$ coordinates we can write
\begin{eqnarray*}
\rho (\mathbf{r}_{1},\mathbf{r}_{2};t) &=&\left( \frac{m\omega }{2\pi \hbar
\left \vert \sin \omega t\right \vert }\right) ^{d}e^{i\frac{m\omega \cos
\omega t}{2\hbar \sin \omega t}(\mathbf{r}_{1}^{2}-\mathbf{r}_{2}^{2})} 
\int \int \rho _{0}\left( \mathbf{u+}\frac{\mathbf{v}}{2},%
\mathbf{u-}\frac{\mathbf{v}}{2}\right) e^{\frac{im\omega }{2\hbar \sin
\omega t}[(2\mathbf{u}\cos \omega t-(\mathbf{r}_{1}+\mathbf{r}_{2})).\mathbf{%
v}-2\mathbf{u.}(\mathbf{r}_{1}-\mathbf{r}_{2})]}\ d\mathbf{u}d\mathbf{v.}
\end{eqnarray*}%
According to the definition in Eq.~\eqref{8}, the Wigner function of this density
matrix can be written as%
\begin{equation}
W(\mathbf{r},\mathbf{p};t) =\left( \frac{m\omega }{2\pi \hbar \left \vert
\sin \omega t\right \vert }\right) ^{d}\int \int e^{\frac{im\omega }{\hbar
\sin \omega t}(\mathbf{u}\cos (\omega t)-\mathbf{r})\mathbf{v}}\rho
_{0}\left( \mathbf{u+}\frac{\mathbf{v}}{2},\mathbf{u-}\frac{\mathbf{v}}{2}%
\right) \,d\mathbf{u}d\mathbf{v}  
\int e^{\frac{i}{\hbar }\left( m\omega \mathbf{r}\cot
g(\omega t)-\frac{m\omega }{\sin \omega t}\mathbf{u}-\mathbf{p}\right) .%
\mathbf{s}}\,d\mathbf{s},  \label{25}
\end{equation}%
with $\mathbf{r}=(\mathbf{r}_{1}+\mathbf{r}_{2})/2$ and $\mathbf{s}=\mathbf{r%
}_{1}-\mathbf{r}_{2}$. We can perform the $d$ dimensional $\mathbf{s}$
integration,%
\begin{eqnarray}
\int e^{\frac{i}{\hbar }\left( m\omega \mathbf{r}\cot g(\omega t)-\frac{%
m\omega }{\sin \omega t}\mathbf{u}-\mathbf{p}\right) .\mathbf{s}}\,d\mathbf{s%
} &=&\left( 2\pi \right) ^{d}\delta \left( \frac{m\omega }{\hbar }\mathbf{r}%
\cot g(\omega t)-\frac{m\omega }{\hbar \sin \omega t}\mathbf{u}-\frac{%
\mathbf{p}}{\hbar }\right)  \notag \\
&=&\left( 2\pi \hbar \right) ^{d}\left( \frac{\left \vert \sin \omega
t\right \vert }{m\omega }\right) ^{d}\delta \left( \mathbf{r}\cos \omega t-%
\mathbf{u}-\frac{\sin \omega t}{m\omega }\mathbf{p}\right)  \label{26}
\end{eqnarray}%
where we have used the property $\delta (a\mathbf{b})=\delta (\mathbf{b}%
)/\left \vert a\right \vert ^{d}$, and therefore  Eq.~\eqref{25} reduces to 
\begin{equation*}
W(\mathbf{r},\mathbf{p};t)=\int \int e^{\frac{im\omega }{\hbar \sin \omega t}%
(\mathbf{u}\cos (\omega t)-\mathbf{r})\mathbf{v}}\rho _{0}\left( \mathbf{u+}%
\frac{\mathbf{v}}{2},\mathbf{u-}\frac{\mathbf{v}}{2}\right) \delta \left( 
\mathbf{r}\cos \omega t-\mathbf{u}-\frac{\sin \omega t}{m\omega }\mathbf{p}%
\right) \,d\mathbf{u}d\mathbf{v}.   
\end{equation*}
After $\mathbf{u}$-integration, we get 
\begin{equation*}
W(\mathbf{r},\mathbf{p};t)=\int e^{-\frac{i}{\hbar }\left[ m\omega \mathbf{r}%
\sin \omega t+\mathbf{p}\cos \omega t\right] .\mathbf{v}}\rho _{0}\left( 
\mathbf{r}\cos \omega t-\frac{\sin \omega t}{m\omega }\mathbf{p+}\frac{%
\mathbf{v}}{2},\mathbf{r}\cos \omega t-\frac{\sin \omega t}{m\omega }\mathbf{%
p-}\frac{\mathbf{v}}{2}\right) \,d\mathbf{v} . 
\end{equation*}%
Notice that according to Eq.~\eqref{8}, the above expression represents the
initial Wigner function at a multidimensional phase space point $\left( 
\mathbf{r}\cos \omega t-\frac{\mathbf{p}\sin \omega t}{m\omega },m\omega 
\mathbf{r}\sin \omega t+\mathbf{p}\cos \omega t\right) $, that is 
\begin{equation}
W(\mathbf{r},\mathbf{p};t)=W_{0}\left( \mathbf{r}\cos \omega t-\frac{\mathbf{%
p}\sin \omega t}{m\omega },m\omega \mathbf{r}\sin \omega t+\mathbf{p}\cos
\omega t\right) ,  \label{29}
\end{equation}%
which ends the proof and reduces to the result given in Eq.~\eqref{yy} for $d=1$.
Notice that if $\omega \rightarrow 0$, this relation reduces to the result
given in Eq. \eqref{15}, as it should.

The existence of alternative derivations for a given problem can enrich one's
insight in solving other problems related to it. In this respect, it is
interesting to see how our generalized phase space result
translates in a real space. In other words, what is the relation between the
one-body density matrix $\rho (\mathbf{r}_{1},\mathbf{r}_{2};t)$ at time $t$
and its initial value $\rho _{0}(\mathbf{r}_{1},\mathbf{r}_{2})$ at $t=0$ in 
$d$ dimensions. To obtain this connection, we shall directly apply the inverse Wigner
transformation to both sides of Eq.~\eqref{9} with \eqref{29}, we obtain the one-body
density matrix in real space at time $t$ 
\begin{equation}
\rho (\mathbf{r}+\mathbf{s}/2,\mathbf{r}-\mathbf{s}/2;t)=\int \frac{d\mathbf{%
p}}{(2\pi \hbar )^{d}}\,W_{0}\left( \mathbf{r}\cos \omega t-\frac{\mathbf{p}%
\sin \omega t}{m\omega },m\omega \mathbf{r}\sin \omega t+\mathbf{p}\cos
\omega t\right) e^{i \mathbf{p \cdot s}/\hbar}.  \label{30}
\end{equation}%
Although the analytical expression of $W_{0}$ in $d=1$  is known 
to be  given in terms of Laguerre polynomials $L_{k}$ for arbitrary $N$  \cite{Hillery} as
$$
W_{0}\left( x,p\right) =2\exp \left\{-\frac{2}{\hbar \omega _{0}}\left( \frac{p^{2}}{%
2m}+\frac{1}{2}m\omega _{0}^{2}x^{2}\right) \sum_{k=0}^{N-1}(-1)^{k}L_{k}\left( \frac{4}{\hbar \omega _{0}}\left( 
\frac{p^{2}}{2m}+\frac{1}{2}m\omega _{0}^{2}{x}^{2}\right) \right) \right\}, 
$$
and its expression in $d$ dimensions can be shown 
to be given in terms of generalized Laguerre polynomials 
$ L_{k}^{(d-1)}$ \cite{Zyl}, 
\begin{equation}\label{generalizedLaguerre}
W_{0}(\mathbf{r},\mathbf{p})=2^{d}   \exp \left\{ -\frac{2}{\hbar \omega
_{0}}\left( \frac{\mathbf{p}^{2}}{2m}+\frac{1}{2}m\omega _{0}^{2}\mathbf{r}%
^{2}\right) \sum_{k=0}^{N-1}(-1)^{k}L_{k}^{(d-1)}\left( \frac{4}{%
\hbar \omega _{0}}\left( \frac{\mathbf{p}^{2}}{2m}+\frac{1}{2}m\omega
_{0}^{2}\mathbf{r}^{2}\right) \right) \right\},
\end{equation}
here its use to calculate the
integral in Eq.~\eqref{30} is not straightforward. 

Alternatively, we propose to
use expression of the so-called Bloch propagator in Wigner phase space which
is related to the density operator through an appropriate Laplace transform.
This relation is obtained as follows. For a system of $N$ \ noninteracting
fermions moving in a potential $V(\mathbf{r})$, the one-body density
matrix is 
$$
\rho _{0}(\mathbf{r}_{1},\mathbf{r}_{2})=\sum_{k}\phi _{k}(
\mathbf{r}_{1})\phi _{k}^{\ast }(\mathbf{r}_{2})\theta (\mu -\varepsilon
_{k}),
$$
where the sum is over occupied single-particle states up to the
Fermi energy $\mu $, and the $\phi _{k}$'s and $\varepsilon _{k}$'s are
respectively the normalized single particle wavefunctions and their
corresponding energies, that is $H_{0}\phi _{k}=\varepsilon _{k}\phi _{k}$. The Heaviside unit-step function is denoted by $\theta (x)$ and by using
the inverse Laplace transform identity \cite{Abramowitz}
$$
\theta (\mu -\varepsilon
_{k})=\int_{c-i\infty }^{c+i\infty }\frac{dz}{2\pi i}\frac{e^{z(\mu
-\varepsilon _{k})}}{z},
$$
we can write the density
matrix $\rho _{0}(\mathbf{r}_{1},\mathbf{r}_{2})$ as an inverse Laplace
transformation (see \cite{Brack} and references quoted therein), so that 
\begin{equation}
\rho _{0}(\mathbf{r}_{1},\mathbf{r}_{2})=\int \limits_{c-i\infty
}^{c+i\infty }\frac{dz}{2\pi i}\ e^{z\mu }\ \frac{C_{0}(\mathbf{r}_{1},\mathbf{r}%
_{2};z)}{z},  \label{31}
\end{equation}%
with $C_{0}(\mathbf{r}_{1},\mathbf{r}_{2};z)=\sum_{k}\phi _{k}(\mathbf{r}%
_{1})\phi _{k}^{\ast }(\mathbf{r}_{2})e^{-z\varepsilon _{k}}$ the matrix
elements of the Bloch operator $e^{-zH_{0}}$. Here the parameter $z$ is
considered as a mathematical variable which in general is taken to be complex
and $c$ is a positive constant. We can immediately obtain the desired
relation by writing the Wigner phase space version of Eq.~\eqref{31} so that 
\begin{equation}
W_{0}(\mathbf{r},\mathbf{p})=\int \limits_{c-i\infty }^{c+i\infty }\frac{dz%
}{2\pi i} \ e^{z\mu } \ \frac{\widetilde{C}_{0}(\mathbf{r},\mathbf{p;}z)}{z},
\label{32}
\end{equation}%
where $\widetilde{C}_{0}(\mathbf{r},\mathbf{p;}z)$ denotes the Wigner
transform of $C_{0}(\mathbf{r}_{1},\mathbf{r}_{2};z)$. For the case of an
isotropic harmonic potential in $d$ dimensions, the phase space function $\widetilde{C}_{0}(%
\mathbf{r},\mathbf{p;}z)$ has a simple explicit expression \cite{Hillery,Ozorio}, 
\begin{equation}
\widetilde{C}_{0}(\mathbf{r},\mathbf{p;}z)=\frac{1}{\left[ \cosh \left( 
\frac{z\hbar \omega _{0}}{2}\right) \right] ^{d}}e^{-\frac{2}{\hbar \omega
_{0}}\left( \tanh \frac{z\hbar \omega _{0}}{2}\right) \left( \frac{\mathbf{p}%
^{2}}{2m}+\frac{1}{2}m\omega _{0}^{2}\mathbf{r}^{2}\right) } . \label{33}
\end{equation}%
To end these preparations, we give the expression of the inverse Wigner
transform of Eq.~\eqref{33}, in terms of centre of mass and relative coordinates (see \cite{BencheikhJPA} and references quoted therein),
\begin{equation}
C_{0}(\mathbf{r}+\mathbf{s}/2,\mathbf{r}-\mathbf{s}/2)=\left( \frac{m\omega 
}{2\pi \hbar \sinh z\hbar \omega }\right) ^{\frac{d}{2}}e^{-\frac{m\omega }{%
\hbar }\left[ \mathbf{r}^{2}\tanh \left( \frac{z\hbar \omega }{2}\right) +%
\frac{\mathbf{s}^{2}}{4}\text{coth}\left( \frac{z\hbar \omega }{2}\right) %
\right] } ,  \label{34}
\end{equation}%
an expression which will be used shortly.

Now we are in position to proceed with the integral in Eq.~\eqref{30}. Let us
substitute Eq.~\eqref{32} into ~\eqref{30}, we obtain%
\begin{equation}
\rho (\mathbf{r}+\tfrac{\mathbf{s}}2,\mathbf{r}-\tfrac{\mathbf{s}}2;t)=\int
\limits_{c-i\infty }^{c+i\infty }\frac{dz}{2\pi i}\frac{e^{z\mu }}{z}\int 
\frac{d\mathbf{p}}{(2\pi \hbar )^{d}}e^{i\frac{\mathbf{p \cdot s}}{\hbar }}%
\widetilde{C}_{0}\left( \mathbf{r}\cos \omega t-\frac{\mathbf{p}\sin \omega t%
}{m\omega },m\omega \mathbf{r}\sin \omega t+\mathbf{p}\cos \omega t;z\right),
\label{35}
\end{equation}%
and using the expression of $\widetilde{C}_{0}$ in Eq.~\eqref{33}, we get 
\begin{eqnarray}
\rho (\mathbf{r}+\mathbf{s}/2,\mathbf{r}-\mathbf{s}/2;t) &=&\int
\limits_{c-i\infty }^{c+i\infty }\frac{dz}{2\pi i}\frac{e^{z\mu }}{z\left[
\cosh \left( \frac{z\hbar \omega _{0}}{2}\right) \right] ^{d}}\int \frac{d%
\mathbf{p}}{(2\pi \hbar )^{d}}e^{i\frac{\mathbf{p \cdot s}}{\hbar }}   \notag
\\
&&\qquad\qquad \times \left[ e^{-\frac{2}{\hbar \omega _{0}}\left( \tanh \frac{z\hbar \omega _{0}%
}{2}\right) \left( \frac{\left( m\omega \mathbf{r}\sin \omega t+\mathbf{p}%
\cos \omega t\right) ^{2}}{2m}+\frac{1}{2}m\omega _{0}^{2}\left( \mathbf{r}%
\cos \omega t-\frac{\mathbf{p}\sin \omega t}{m\omega }\right) ^{2}\right) }%
\right] . \label{36}
\end{eqnarray}%
The above integral on $\mathbf{p}$ is carried out in the Appendix B, and therefore we have
\begin{equation}
\rho (\mathbf{r}+\tfrac{\mathbf{s}}2,\mathbf{r}-\tfrac{\mathbf{s}}2;t)=\frac{e^{i\frac{m}{%
\hbar }\frac{\dot{b}(t)}{b(t)}\mathbf{r} \cdot \mathbf{s}}}{ (b(t))^{d}}\int
\limits_{c-i\infty }^{c+i\infty }\frac{dz}{2\pi i}\frac{e^{z\mu }}{z} 
\left( \frac{m\omega _{0}\text{ }}{2\pi \sinh z\hbar \omega _{0}}\right) ^{%
\frac{d}{2}}e^{-\frac{m\omega }{\hbar }\left[ \frac{\mathbf{r}^{2}}{b^{2}}%
\tanh \left( \frac{z\hbar \omega }{2}\right) +\frac{\mathbf{s}^{2}}{4b^{2}}%
\text{coth}\left( \frac{z\hbar \omega }{2}\right) \right] }, 
\label{37}
\end{equation}%
where $b(t)$ is a time dependent scaling factor given by 
\begin{equation}
b(t)=\sqrt{1+\left( \frac{\omega _{0}^{2}}{\omega ^{2}}-1\right) \sin
^{2}\omega t} , \label{38}
\end{equation}%
which is the solution of $\dot{b}(t)=\left( \frac{\omega _{0}^{2}}{\omega ^{2}}%
-1\right) \left( \omega \sin \omega t\cos \omega t\right) /b(t)$. Equation \eqref{34} allows us to write Eq.~{\eqref{37} as 
\begin{equation}
\rho (\mathbf{r}+\mathbf{s}/2,\mathbf{r}-\mathbf{s}/2;t)=\frac{e^{i\frac{m}{%
\hbar }\frac{\dot{b}(t)}{b(t)}\mathbf{r} \cdot \mathbf{s}}}{(b(t))^{d}}\int
\limits_{c-i\infty }^{c+i\infty }\frac{dz}{2\pi i} \ \frac{e^{z\mu }}{z} \ 
C_{0}\left( \frac{\mathbf{r}}{b(t)}+\frac{\mathbf{s}}{2b(t)},\frac{\mathbf{r}}{b(t)}-%
\frac{\mathbf{s}}{2b(t)}\right) . \label{39}
\end{equation}%
We observe that the above inverse Laplace transform is nothing but the
initial one-body density matrix at rescaled positions, with $b(t)$ as the
scaling factor. We then arrive to 
\begin{equation}
\rho (\mathbf{r}+\mathbf{s}/2,\mathbf{r}-\mathbf{s}/2;t)=\frac{e^{i\frac{m}{%
\hbar }\frac{\dot{b}(t)}{b(t)}\mathbf{r} \cdot \mathbf{s}}}{(b(t))^{d}} \ \rho
_{0}\left( \frac{\mathbf{r}}{b(t)}+\frac{\mathbf{s}}{2b(t)},\frac{\mathbf{r}}{b(t)}-%
\frac{\mathbf{s}}{2b(t)}\right) , \label{40}
\end{equation}%
returning to the original coordinates, $\mathbf{r}_{1}=$ $\mathbf{r}+\mathbf{%
s}/2$, $\mathbf{r}_{2}=$ $\mathbf{r}-\mathbf{s}/2$, and since $\mathbf{r} \cdot
\mathbf{s=(r}_{1}^{2}-\mathbf{r}_{2}^{2})/2$, the above scaling law becomes 
\begin{equation}
\rho (\mathbf{r}_{1},\mathbf{r}_{2};t)=\frac{e^{i\frac{m}{2\hbar }\frac{%
\dot{b}(t)}{b(t)}\mathbf{(r}_{1}^{2}-\mathbf{r}_{2}^{2})}}{(b(t))^{d}} \ 
\rho_{0}\left( \frac{\mathbf{r}_{1}}{b(t)},\frac{\mathbf{r}_{2}}{b(t)}\right),
\label{41}
\end{equation}%
a result valid for arbitrary dimensions. For $d=1$, one recovers the
result obtained by the rescaling method \cite{DeanEPL}.

\subsection{Ballistic versus non-ballistic expansions in phase space}

It is important to note that the above scaling law obtained for trap to trap
quench is not restricted just to noninteracting particles. In fact, the
scaling law in Eq.~\eqref{41} was proven to hold also for some physical
systems of interacting particles in which the interactions are acting
before and after the quench of the harmonic potential (non ballistic
expansion). This is the case for the two following situations: (i) the Tonks-Girardeau
gas, which consists in a gas of identical bosons interacting through very
strong repulsive zero-range interactions, confined by harmonic trap in one
dimension $(d=1)$ \cite{Minguzzi,Ruggiero}, and (ii) for harmonically trapped
interacting fermions in three dimensions $(d=3)$ at unitarity  \cite{Castin}.
Notice that long time ago Pitaevskii and Rosch \cite{PitaevskiiRosch} 
introduced a scaling ansatz for a two dimensional bosonic system of particles interacting
with contact or inverse square interaction. Later on, scaling approach to
quantum non-equilibrium dynamics of interacting systems subject to external
linear and parabolic potentials has been examined in \cite{Gritsev}, 
where many-body scaling solutions to more general types of interaction and
arbitrary dimensionality where obtained.

It may be of interest to see how the above scaling law is expressed
in Wigner phase space. For obtaining this result,  we substitute Eq.~\eqref{40} into ~\eqref{8} 
and we obtain 
\begin{eqnarray}
W(\mathbf{r},\mathbf{p};t) &=&\frac{1}{(b(t))^d}\int d\mathbf{s}\ \rho
_{0}\left( \frac{\mathbf{r}}{b(t)}+\frac{\mathbf{s}}{2b(t)},\frac{\mathbf{r}}{b(t)}-%
\frac{\mathbf{s}}{2b(t)}\right) \,e^{-\frac{i}{\hbar }\left( b(t)\mathbf{p-}m%
\dot{b}(t)\mathbf{r}\right) \cdot \frac{\mathbf{s}}{b(t)}}  
\notag \\ &=&
\int d\mathbf{u}\ \rho _{0}\left( \frac{\mathbf{r}}{b(t)}+\frac{\mathbf{u}}{2%
},\frac{\mathbf{r}}{b(t)}-\frac{\mathbf{u}}{2}\right) \,e^{-\frac{i}{\hbar }%
\left( b(t)\mathbf{p-}m\dot{b}(t)\mathbf{r}\right) \cdot \mathbf{\mathbf{u}}},
\label{42}
\end{eqnarray}%
and according to Eq.~\eqref{8}, the right-hand side represent the initial Wigner
function at phase space point $\left( \frac{\mathbf{r}}{b(t)},b(t)\mathbf{p-}m%
\dot{b}(t)\mathbf{r}\right) $, so that 
\begin{equation}
W(\mathbf{r},\mathbf{p};t)=W_{0}\left( \frac{\mathbf{r}}{b(t)},b(t)\mathbf{p-}m%
\dot{b}(t)\mathbf{r}\right) , 
\label{43}
\end{equation}%
where we recall that $b(t)$ is given by \eqref{38}. 

As stated for Eq.~\eqref{41}, it follows that the
relation in Eq.~\eqref{43} is not only valid for noninteracting particle systems
but also for the above two systems pertaining to interacting particles. The
interested reader may ask on the difference between Wigner functions given
respectively by Eq.~\eqref{29} and \eqref{43}. The time dependent Wigner function in Eq.~\eqref{29} describes a ballistic expansion (interactions are suppressed) following
the quench of the harmonic potential while the Wigner function in Eq.~\eqref{43}
concerns ballistic or a nonballistic expansion after the quench of the
harmonic trap pertaining to the nature (interacting or noninteracting) of
the initially confined system before the quench. We can show that for an
initially harmonically confined noninteracting particle system subjected to
a quench of the potential, the two forms Eq.~\eqref{29} and \eqref{43} are identical,
that is 
\begin{equation}
W_{0}\left( \mathbf{r}\cos \omega t-\frac{\mathbf{p}\sin \omega t}{m\omega }%
,m\omega \mathbf{r}\sin \omega t+\mathbf{p}\cos \omega t\right) =W_{0}\left( 
\frac{\mathbf{r}}{b(t)},b\mathbf{p-}m\dot{b}(t) \mathbf{r}\right) .
  \label{44}
\end{equation}
In fact for a system of $N$ noninteracting fermions confined in a $d
$ dimensional isotropic harmonic potential with frequency $\omega _{0}$, the
Wigner function was given in \eqref{generalizedLaguerre}, 
where we can observe that $W_{0}(\mathbf{r},\mathbf{p})$ depends on the phase space
variables $(\mathbf{r},\mathbf{p})$
by means of the classical Hamiltonian $H_{cl}(\mathbf{r},\mathbf{p})=
\left( \frac{\mathbf{p}^{2}}{2m}+\frac{1}{2}m\omega _{0}^{2}\mathbf{r}^{2}\right) $, 
and is nothing but the Wigner transform of the quantum one
particle Hamiltonian. Hence, to prove Eq.~\eqref{44} one has just
to check the equality 
\begin{equation*}
\frac{\left( m\omega \mathbf{r}\sin \omega t+\mathbf{p}\cos \omega t\right) ^{2}}{2m} 
+\frac{m\omega ^{2}}{2}\left( \mathbf{r}\cos \omega t-\frac{\mathbf{p}\sin \omega t}{m\omega }\right) ^{2}=\frac{\left( b(t) \mathbf{p-}m\dot{b}(t)\mathbf{r}\right) ^{2}}{2m}+\frac{m\omega^{2}}{2} \left( \frac{\mathbf{r}}{b(t)}\right) ^{2} . 
\end{equation*}
Recalling \eqref{38} it is easy to verify the last equation.

We believe that for interacting particles Eq.~\eqref{44} is no longer true, therefore in that case
$$
W_{0}\left( \mathbf{r}\cos \omega t-\frac{\mathbf{p}\sin \omega t}{%
m\omega },m\omega \mathbf{r}\sin \omega t+\mathbf{p}\cos \omega t\right)
\neq W_{0}\left( \frac{\mathbf{r}}{b(t)},b(t) \mathbf{p-}m\dot{b}(t)
\mathbf{r}\right).
$$
As stated before, Eq.~\eqref{29} is valid for both interacting or noninteracting
system of particles providing that the gas expands ballistically
(suppression of interactions) after the quench. In this respect we shall
exploit this relation to see how one can access to the initial momentum
density of the interacting system using the expansion of the density in real
space.

\subsection{Recovering the initial momentum density}

Recently, an experimental technique to directly image the
momentum distribution of a strongly interacting two-dimensional quantum gas
was obtained and characterized  \cite{Murthy}. This method is based on the fact that, just after 
switching-off the initially confining trap, and instead of a free expansion, the
gas is subjected to an external harmonic potential $V(\mathbf{r})=m\omega
^{2}\mathbf{r}^{2}/2$ where the gas moves ballistically (the interactions
are suppressed). It was shown that after a quarter of the oscillator time
period $T=2\pi /\omega $, the spatial distribution is related to the momentum
density of the initially confined quantum gas. In the following, we provide a
generalization of this relation in arbitrary spatial dimensions by exploiting the relation in Eq.~\eqref{29} for the Wigner function.

Let us denote by $T$ the period corresponding to frequency $\omega $ of the
harmonic trap and considering the specific time $t=T/4=\pi /2\omega $ after
the quench, Eq.~\eqref{29} reduces to
\begin{equation*}
W\left( \mathbf{r},\mathbf{p};\frac{T}{4}\right) =W_{0}\left( -\frac{\mathbf{%
p}}{m\omega },m\omega \mathbf{r}\right) .
\end{equation*}%
Using Eq.~\eqref{10}, the spatial density  at this time is
\begin{equation*}
\rho \left( \mathbf{r};\frac{T}{4}\right) =\int \frac{d\mathbf{p}}{(2\pi
\hbar )^{d}}\,W_{0}\left( -\frac{\mathbf{p}}{m\omega },m\omega \mathbf{r}%
\right) .
\end{equation*}%
Making the change of variable $\mathbf{u=-p/(}m\omega )$, the above integral
in $d$ dimensions becomes%
\begin{equation}
\rho \left( \mathbf{r};\frac{T}{4}\right) =\mathbf{(}m\omega )^{d}
\int \frac{d\mathbf{u}}{(2\pi \hbar )^{d}}
\,W_{0}\left( \mathbf{u},m\omega \mathbf{r}%
\right) , \label{xx}
\end{equation}%
and according to Eq.~\eqref{11}, the right-hand side of Eq.~\eqref{xx} essentially represents the
initial momentum density, that is 
\begin{equation}
\rho \left( \mathbf{r};\frac{T}{4}\right) =\mathbf{(}m\omega )^{d}n(\mathbf{%
p=}m\omega \mathbf{r};0) , \label{48}
\end{equation}%
a relation which clearly exhibits the mapping between the initial
momentum density for a given value of the momentum $\mathbf{p=}m\omega 
\mathbf{r}$ and the spatial density at position $\mathbf{r}$ at time $t=T/4$
after the ballistic expansion in the harmonic trap with frequency $\omega $.

\section{Concluding remarks}
\label{remarks}

In this paper we have studied the non-equilibrium dynamics in phase space
generated by a sudden change of the Hamiltonian in a quantum system, through
the analysis of the Wigner function. For the case of two attractive
particles we calculated the corresponding time dependent Wigner function
following a swich-off of the interaction.

For possible experimental implementation in ultra-cold quantum gases field,
an interaction of the form $-a\delta (x)+b\delta ^{\prime }(x),\ a>0,\ b\in 
\mathbb{R}$ was considered and we have calculated the two-particle Wigner
function. At large values of momentum $p$, we have found that the associated
momentum distribution scales as $1/p^{2}$, while in the absence of $\delta
^{\prime }$ interaction, this density scales, in this case of pure
zero-range delta interaction, as the well known law $\ 1/p^{4}$.

We have generalized to arbitrary dimensions $d$ a derivation, by using
alternative method, of a relationship shown recently in one dimension
between the Wigner function at time $t$ and its initial value following a
sudden change of the harmonic trap of noninteracting particles. We have
exploited our generalized relation, through the use of inverse Wigner
transformation to obtain in $d$ dimensions the scaling law satisfied by the
one-body density matrix in real space. Using the generalized
relation in Wigner phase space for the considered quench, we have shown
that the initial momentum density of a system of particles (interacting or
noninteracting) is exactly mapped for a given value of the momentum 
$\mathbf{p=}m\omega \mathbf{r}$ to the spatial density at position $\mathbf{r}$ at
time a quarter of time period, $t=T/4$, after the ballistic expansion in the
harmonic trap with frequency $\omega $. 
It should be noted that our method can be easily adapted to deal with physical
situations corresponding to quasi-$1d$ or quasi-$2d$ configurations.

An interesting extension of this dynamically situation would be to study the
problem of a Lieb-Liniger gas at finite repulsion strength. Work in this
direction is in progress.

\section*{Acknowledgements}

This work was partially supported by the Spanish MINECO
(MTM2014-57129-C2-1-P), Junta de Castilla y Le\'on and FEDER projects
(BU229P18, VA057U16, and VA137G18). K.B. thanks the Direction G\'en\'erale de la Recherche Scientifique et du D\'eveloppement Technologique (DGRSDT-Algeria) for financial support. The authors acknowledge the anonymous referee for helpful suggestions.

\section*{Appendix A}

In this Appendix we shall prove the relationship given in Eq. \eqref{6new}.
Using Dirac notations, let us rewrite Eq. \eqref{5new} as 
\begin{eqnarray}
\rho (\mathbf{r},\mathbf{r}^{\prime };t) &=&N\int d\mathbf{r}_{2}\cdots d%
\mathbf{r}_{N}\left \langle \mathbf{r},\mathbf{r}_{2},\dots ,\mathbf{r}%
_{N}\right \vert \left. \Phi (t)\right \rangle \left \langle \Phi (t)\right
\vert \left. \mathbf{r}^{\prime },\mathbf{r}_{2},\dots ,\mathbf{r}_{N}\right
\rangle  \notag \\
&=&N\int d\mathbf{r}_{2}\cdots d\mathbf{r}_{N}\left \langle \mathbf{r},%
\mathbf{r}_{2},\dots ,\mathbf{r}_{N}\right \vert e^{-\frac{i}{\hbar }%
Ht}\left \vert \Phi _{0}\right \rangle \left \langle \Phi _{0}\right \vert
e^{+\frac{i}{\hbar }Ht}\left \vert \mathbf{r}^{\prime },\mathbf{r}_{2},\dots
,\mathbf{r}_{N}\right \rangle ,  \label{A2}
\end{eqnarray}%
where we have used Eq. \eqref{4new} to obtain the second form. Applying the
closure relationship, we can write 
\begin{eqnarray}
\rho (\mathbf{r},\mathbf{r}^{\prime };t) &=&N\int d\mathbf{r}_{2}\cdots d%
\mathbf{r}_{N}\int d\mathbf{r}_{1}^{\prime }\cdots d\mathbf{r}_{N}^{\prime
}\int d\mathbf{r}_{1}^{\prime \prime }\cdots d\mathbf{r}_{N}^{\prime \prime
}\ \Phi _{0}(\mathbf{r}_{1}^{\prime },\dots ,\mathbf{r}_{N}^{\prime })\Phi
_{0}^{\ast }(\mathbf{r}_{1}^{\prime \prime },\dots ,\mathbf{r}_{N}^{\prime
\prime })  \label{A3} \\
&&\qquad\qquad \times \left \langle \mathbf{r},\mathbf{r}_{2},\dots ,\mathbf{r}%
_{N}\right \vert e^{-\frac{i}{\hbar }Ht}\left \vert \mathbf{r}_{1}^{\prime },%
\mathbf{r}_{2}^{\prime },\dots ,\mathbf{r}_{N}^{\prime }\right \rangle \left
\langle \mathbf{r}_{1}^{\prime \prime },\mathbf{r}_{2}^{\prime \prime
},\dots ,\mathbf{r}_{N}^{\prime \prime }\right \vert e^{+\frac{i}{\hbar }%
Ht}\left \vert \mathbf{r}^{\prime },\mathbf{r}_{2},\mathbf{r}_{3},\dots ,%
\mathbf{r}_{N}\right \rangle .  \notag
\end{eqnarray}%
Since the post-quench Hamiltonian $H$ describes a system of $N$
noninteracting particles, we can write the following factorization for the
time evolution operator, so that 
\begin{equation}
e^{-\frac{i}{\hbar }Ht}=e^{-\frac{it}{\hbar }\left( \frac{\mathbf{p}_{1}^{2}%
}{2m}+V(\mathbf{r}_{1})\right) }\ e^{-\frac{it}{\hbar }\left( \frac{\mathbf{p%
}_{2}^{2}}{2m}+V(\mathbf{r}_{2})\right) }\cdots e^{-\frac{it}{\hbar }\left( 
\frac{\mathbf{p}_{N}^{2}}{2m}+V(\mathbf{r}_{N})\right) },  \label{A4}
\end{equation}%
from which we immediately deduce 
\begin{equation*}
\left \langle \mathbf{r},\mathbf{r}_{2},\cdots ,\mathbf{r}_{N}\right \vert
e^{-\frac{it}{\hbar }H}\left \vert \mathbf{r}_{1}^{\prime },\mathbf{r}%
_{2}^{\prime },\dots ,\mathbf{r}_{N}^{\prime }\right \rangle =\left \langle 
\mathbf{r}\right \vert e^{-\frac{it}{\hbar }\left( \frac{\mathbf{p}_{1}^{2}}{%
2m}+V(\mathbf{r}_{1})\right) }\left \vert \mathbf{r}_{1}^{\prime }\right
\rangle \cdots \left \langle \mathbf{r}_{N}\right \vert e^{-\frac{it}{\hbar }%
\left( \frac{\mathbf{p}_{N}^{2}}{2m}+V(\mathbf{r}_{N})\right) }\left \vert 
\mathbf{r}_{N}^{\prime }\right \rangle ,
\end{equation*}%
and similarly 
\begin{equation*}
\left \langle \mathbf{r}_{1}^{\prime \prime },\mathbf{r}_{2}^{\prime \prime
},\dots ,\mathbf{r}_{N}^{\prime \prime }\right \vert e^{\frac{it}{\hbar }%
H}\left \vert \mathbf{r}^{\prime },\mathbf{r}_{2},\mathbf{r}_{3},\dots ,%
\mathbf{r}_{N}\right \rangle =\left \langle \mathbf{r}_{1}^{\prime \prime
}\right \vert e^{\frac{it}{\hbar }\left( \frac{\mathbf{p}_{1}^{2}}{2m}+V(%
\mathbf{r}_{1})\right) }\left \vert \mathbf{r}^{\prime }\right \rangle
\cdots \left \langle \mathbf{r}_{N}^{\prime \prime }\right \vert e^{\frac{it%
}{\hbar }\left( \frac{\mathbf{p}_{N}^{2}}{2m}+V(\mathbf{r}_{N})\right)
}\left \vert \mathbf{r}_{N}\right \rangle .
\end{equation*}%
Substituting these last two relations into Eq. \eqref{A3} we obtain 
\begin{eqnarray*}
\rho (\mathbf{r},\mathbf{r}^{\prime };t) &=&N\int d\mathbf{r}_{2}\cdots d%
\mathbf{r}_{N}\int d\mathbf{r}_{1}^{\prime }\cdots d\mathbf{r}_{N}^{\prime
}\int d\mathbf{r}_{1}^{\prime \prime }\cdots d\mathbf{r}_{N}^{\prime \prime
}\ \Phi _{0}(\mathbf{r}_{1}^{\prime },\dots ,\mathbf{r}_{N}^{\prime })\Phi
_{0}^{\ast }(\mathbf{r}_{1}^{\prime \prime },,\dots ,\mathbf{r}_{N}^{\prime
\prime }) \\
&&\qquad\times \left \langle \mathbf{r}\right \vert e^{-\frac{it}{\hbar }\left( 
\frac{\mathbf{p}_{1}^{2}}{2m}+V(\mathbf{r}_{1})\right) }\left \vert \mathbf{r%
}_{1}^{\prime }\right \rangle \left \langle \mathbf{r}_{2}\right \vert e^{-%
\frac{it}{\hbar }\left( \frac{\mathbf{p}_{2}^{2}}{2m}+V(\mathbf{r}%
_{2})\right) }\left \vert \mathbf{r}_{2}^{\prime }\right \rangle \cdots
\left \langle \mathbf{r}_{N}\right \vert e^{-\frac{it}{\hbar }\left( \frac{%
\mathbf{p}_{N}^{2}}{2m}+V(\mathbf{r}_{N})\right) }\left \vert \mathbf{r}%
_{N}^{\prime }\right \rangle \\
&&\qquad\times \left \langle \mathbf{r}_{1}^{\prime \prime }\right \vert e^{\frac{%
it}{\hbar }\left( \frac{\mathbf{p}_{1}^{2}}{2m}+V(\mathbf{r}_{1})\right)
}\left \vert \mathbf{r}^{\prime }\right \rangle \left \langle \mathbf{r}%
_{2}^{\prime \prime }\right \vert e^{\frac{it}{\hbar }\left( \frac{\mathbf{p}%
_{2}^{2}}{2m}+V(\mathbf{r}_{2})\right) }\left \vert \mathbf{r}_{2}\right
\rangle \cdots \left \langle \mathbf{r}_{N}^{\prime \prime }\right \vert e^{%
\frac{it}{\hbar }\left( \frac{\mathbf{p}_{N}^{2}}{2m}+V(\mathbf{r}%
_{N})\right) }\left \vert \mathbf{r}_{N}\right \rangle ,
\end{eqnarray*}%
which can be rewritten as follows 
\begin{eqnarray}
\rho (\mathbf{r},\mathbf{r}^{\prime };t) &=&N\int d\mathbf{r}_{1}^{\prime
}\cdots d\mathbf{r}_{N}^{\prime }\int d\mathbf{r}_{1}^{\prime \prime }\cdots
d\mathbf{r}_{N}^{\prime \prime }\ \Phi _{0}(\mathbf{r}_{1}^{\prime },\dots ,%
\mathbf{r}_{N}^{\prime })\Phi _{0}^{\ast }(\mathbf{r}_{1}^{\prime \prime
},\dots ,\mathbf{r}_{N}^{\prime \prime })  \notag \\
&&\qquad   \times \left \langle \mathbf{r}\right \vert
e^{-\frac{it}{\hbar }\left( \frac{\mathbf{p}_{1}^{2}}{2m}+V(\mathbf{r}%
_{1})\right) }\left \vert \mathbf{r}_{1}^{\prime }\right \rangle \left
\langle \mathbf{r}_{1}^{\prime \prime }\right \vert e^{\frac{it}{\hbar }%
\left( \frac{\mathbf{p}_{1}^{2}}{2m}+V(\mathbf{r}_{1})\right) }\left \vert 
\mathbf{r}^{\prime }\right \rangle  \notag \\
&&\quad \int d\mathbf{r}_{2}\left \langle \mathbf{r}_{2}^{\prime \prime
}\right \vert e^{\frac{it}{\hbar }\left( \frac{\mathbf{p}_{2}^{2}}{2m}+V(%
\mathbf{r}_{2})\right) }\left \vert \mathbf{r}_{2}\right \rangle \left
\langle \mathbf{r}_{2}\right \vert e^{-\frac{it}{\hbar }\left( \frac{\mathbf{%
p}_{2}^{2}}{2m}+V(\mathbf{r}_{2})\right) }\left \vert \mathbf{r}_{2}^{\prime
}\right \rangle \times \cdots  \notag \\
&&\quad \int d\mathbf{r}_{N}\left \langle \mathbf{r}_{N}^{\prime \prime
}\right \vert e^{\frac{it}{\hbar }\left( \frac{\mathbf{p}_{N}^{2}}{2m}+V(%
\mathbf{r}_{N})\right) }\left \vert \mathbf{r}_{N}\right \rangle \left
\langle \mathbf{r}_{N}\right \vert e^{-\frac{it}{\hbar }\left( \frac{\mathbf{%
p}_{N}^{2}}{2m}+V(\mathbf{r}_{N})\right) }\left \vert \mathbf{r}_{N}^{\prime
}\right \rangle .  \label{A8}
\end{eqnarray}%
Using the closure relations over the kets $\left \vert \mathbf{r}%
_{2}\right
\rangle ,\left \vert \mathbf{r}_{3}\right \rangle \dots
.\left
\vert \mathbf{r}_{N}\right \rangle $ and then carrying out the
integrals over the variables $\mathbf{r}_{2}^{\prime \prime },\dots ,\mathbf{%
r}_{N}^{\prime \prime }$, the above expression reduces to 
\begin{eqnarray}
\rho (\mathbf{r},\mathbf{r}^{\prime };t) &=&N\int \int d\mathbf{r}%
_{1}^{\prime }d\mathbf{r}_{1}^{\prime \prime }\left \langle \mathbf{r}\right
\vert e^{-\frac{it}{\hbar }\left( \frac{\mathbf{p}_{1}^{2}}{2m}+V(\mathbf{r}%
_{1})\right) }\left \vert \mathbf{r}_{1}^{\prime }\right \rangle \left
\langle \mathbf{r}_{1}^{\prime \prime }\right \vert e^{\frac{it}{\hbar }%
\left( \frac{\mathbf{p}_{1}^{2}}{2m}+V(\mathbf{r}_{1})\right) }\left \vert 
\mathbf{r}^{\prime }\right \rangle  \notag \\
&&\qquad \times \int d\mathbf{r}_{2}^{\prime }\cdots d\mathbf{r}_{N}^{\prime
}\ \Phi _{0}(\mathbf{r}_{1}^{\prime },\mathbf{r}_{2}^{\prime },\dots ,%
\mathbf{r}_{N}^{\prime })\Phi _{0}^{\ast }(\mathbf{r}_{1}^{\prime \prime },%
\mathbf{r}_{2}^{\prime },\dots ,\mathbf{r}_{N}^{\prime }).  \label{A10}
\end{eqnarray}%
Taking into account the definition of the initial reduced one-body density
matrix in Eq. \eqref{2new}, we arrive to 
\begin{equation}
\rho (\mathbf{r},\mathbf{r}^{\prime };t)=N\int \int d\mathbf{r}_{1}^{\prime
}d\mathbf{r}_{1}^{\prime \prime }\left \langle \mathbf{r}\right \vert e^{-%
\frac{it}{\hbar }\left( \frac{\mathbf{p}_{1}^{2}}{2m}+V(\mathbf{r}%
_{1})\right) }\left \vert \mathbf{r}_{1}^{\prime }\right \rangle \left
\langle \mathbf{r}_{1}^{\prime \prime }\right \vert e^{\frac{it}{\hbar }%
\left( \frac{\mathbf{p}_{1}^{2}}{2m}+V(\mathbf{r}_{1})\right) }\left \vert 
\mathbf{r}^{\prime }\right \rangle \rho _{0}(\mathbf{r}_{1}^{\prime },%
\mathbf{r}_{1}^{\prime \prime })  \label{A11}
\end{equation}%
which we rewrite as 
\begin{equation}
\rho (\mathbf{r},\mathbf{r}^{\prime };t)=\int \int d\mathbf{r}_{1}^{\prime }d%
\mathbf{r}_{1}^{\prime \prime }\ U(\mathbf{r},\mathbf{r}_{1}^{\prime };t)\
U^{\ast }(\mathbf{r}^{\prime },\mathbf{r}_{1}^{\prime \prime };t)\ \rho _{0}(%
\mathbf{r}_{1}^{\prime },\mathbf{r}_{1}^{\prime \prime }),  \label{A12}
\end{equation}%
where $U(\mathbf{r},\mathbf{r}_{1}^{\prime };t)=\left \langle \mathbf{r}%
\right \vert $ $e^{-\frac{it}{\hbar }(\frac{\mathbf{p}_{1}^{2}}{2m}+V(%
\mathbf{r}_{1}))}\left \vert \mathbf{r}_{1}^{\prime }\right \rangle $ is the
single particle propagator associated to the many-body noninteracting
post-quench Hamiltonian. Now changing the names of the variables in Eq. %
\eqref{A12}, so that, $\mathbf{r}_{1}^{\prime }=\mathbf{\xi }_{1}$, $\mathbf{r}%
_{1}^{\prime \prime }=\mathbf{\xi }_{2}$, the desired result in Eq. \eqref{6new}
is recovered.

\section*{Appendix B. Derivation of Eq.~\eqref{37}}

To calculate the integral on $\mathbf{p}$ in Eq.~\eqref{36}, we need some
intermediate results. First of all, it is easy to check that
\begin{eqnarray}
&&\frac{\left( m\omega \mathbf{r}\sin \omega t+\mathbf{p}\cos \omega
t\right) ^{2}}{2m}+\frac{1}{2}m\omega _{0}^{2}\left( \mathbf{r}\cos \omega t-%
\frac{\mathbf{p}\sin \omega t}{m\omega }\right) ^{2} = \\  \notag
&& \qquad\qquad = \frac{1}{2}%
m\omega ^{2}\left[ 1+\frac{\omega _{0}^{2}}{\omega ^{2}}-b^{2}(t)\right] 
\mathbf{r}^{2}+   
\frac{1}{2m}b^{2}(t)\mathbf{p}^{2}-\omega \left( \frac{\omega _{0}^{2}}{%
\omega ^{2}}-1\right) \left( \sin \omega t\cos \omega t\right) \mathbf{r}\cdot 
\mathbf{p}  \label{B1}
\end{eqnarray}%
where $b(t)=\sqrt{1+\left( \frac{\omega _{0}^{2}}{\omega ^{2}}-1\right) \sin
^{2}\omega t}$ is a time dependent scaling factor. After insertion of this result in Eq.~\eqref{36}, it
becomes 
\begin{eqnarray}
\rho (\mathbf{r}+\mathbf{s}/2,\mathbf{r}-\mathbf{s}/2;t) &=&\int
\limits_{c-i\infty }^{c+i\infty }\frac{dz}{2\pi i}\frac{e^{z\mu }}{z\left[
\cosh \left( \frac{z\hbar \omega _{0}}{2}\right) \right] ^{d}}e^{-\frac{%
m\omega ^{2}}{\hbar \omega _{0}}\left( \tanh \frac{z\hbar \omega _{0}}{2}%
\right) \left[ 1+\frac{\omega _{0}^{2}}{\omega ^{2}}-b^{2}(t)\right] \mathbf{%
r}^{2}}  \label{B2}   \\
&&\qquad \times \int \frac{d\mathbf{p}}{(2\pi \hbar )^{d}}e^{i\frac{\mathbf{p \cdot s}}{\hbar }%
}e^{-\frac{2}{\hbar \omega _{0}}\left( \tanh \frac{z\hbar \omega _{0}}{2}%
\right) \left( \frac{1}{2m}b^{2}(t)\mathbf{p}^{2}-\omega \left( \frac{\omega
_{0}^{2}}{\omega ^{2}}-1\right) \left( \sin \omega t\cos \omega t\right) 
\mathbf{r} \cdot \mathbf{p}\right) }  .  \notag
\end{eqnarray}%
Using the $d$-dimensional identity
\begin{equation}
\int d\mathbf{u}\ e^{-\alpha \mathbf{u}^{2}}e^{-\mathbf{a \cdot u}}=\left( \frac{\pi 
}{\alpha }\right) ^{\frac{d}{2}}e^{\frac{\mathbf{a}^{2}}{4\alpha }},
\label{B3}
\end{equation}%
the integration on $\mathbf{p}$ in Eq.~\eqref{B2} can then be performed, yielding
to the expression of the density matrix in the form
\begin{eqnarray*}
&&\rho (\mathbf{r}+\tfrac{\mathbf{s}}2,\mathbf{r}-\tfrac{\mathbf{s}}2;t) =
\frac{1}{(2\pi \hbar \, b(t))^d}\int\limits_{c-i\infty }^{c+i\infty }\frac{dz}{2\pi i%
}\left[ \frac{\pi \hbar m\omega _{0}}{\tanh \frac{z\hbar \omega _{0}}{2}}%
\right]^{\frac{d}{2}}\frac{e^{z\mu }e^{-\frac{m\omega ^{2}}{\hbar \omega
_{0}}\left( \tanh \frac{z\hbar \omega _{0}}{2}\right) \left[ 1+\frac{\omega
_{0}^{2}}{\omega ^{2}}-(b(t))^{2}(t)\right] \mathbf{r}^{2}}}{z\left[ \cosh \left( 
\frac{z\hbar \omega _{0}}{2}\right) \right] ^{d}}   \notag \\
&& \hskip3cm  \times e^{\frac{m\omega ^{2}}{\hbar \omega _{0}(b(t))^{2}}\left( \tanh \frac{%
z\hbar \omega _{0}}{2}\right) \left( \frac{\omega _{0}^{2}}{\omega ^{2}}%
-1\right) ^{2}\left( \sin \omega t\cos \omega t\right) ^{2}\mathbf{r}^{2}-%
\frac{m\omega _{0}}{4\hbar (b(t))^{2}}\left( \coth \frac{z\hbar \omega _{0}}{2}%
\right) \mathbf{s}^{2}+i\frac{m\omega }{\hbar _{0}(b(t))^{2}}\left( 
\frac{\omega_{0}^{2}}{\omega ^{2}}-1\right) \left( \sin \omega t\cos \omega t\right) 
\mathbf{r}\cdot \mathbf{s} },  
\end{eqnarray*}
which after collecting similar terms and simplification turns into Eq.~\eqref{37}.

\end{document}